\documentclass[aps,prc,twocolumn,endfloats,floatfix,showpacs,amsmath,amssymb]{revtex4}
\usepackage{graphicx}
\usepackage{dcolumn}
\usepackage{color} 
\begin{document}
\preprint{UCRL-JRNL-217766}
\title{$^7$Be(p,$\gamma$)$^8$B S-factor from \emph{ab initio} no-core shell model wave functions}
\author{P. Navr\'atil}
\email[]{navratil1@llnl.gov}
\affiliation{Lawrence Livermore National Laboratory, P.O. Box 808, L-414,
Livermore, CA  94551, USA}
\author{C. A. Bertulani}
\email[]{bertulani@physics.arizona.edu}
\affiliation{Department of Physics, University of Arizona, Tucson, AZ 85721, USA}
\author{E. Caurier}
\email[]{etienne.caurier@ires.in2p3.fr}
\affiliation{Institut de Recherches Subatomiques
            (IN2P3-CNRS-Universit\'e Louis Pasteur)\\
            Batiment 27/1,
            67037 Strasbourg Cedex 2, France}

\date{\today}

\begin{abstract}
Nuclear structure of $^7$Be, $^8$B and $^{7,8}$Li is studied within
the {\it ab initio} no-core shell model (NCSM). Starting from
high-precision nucleon-nucleon (NN) interactions, wave functions of
$^7$Be and $^8$B bound states are obtained in basis spaces up to
$10\hbar\Omega$ and used to calculate channel cluster form factors
(overlap integrals) of the $^8$B ground state with $^7$Be+p.
Due to the use of the harmonic oscillator (HO) basis, 
the overlap integrals have incorrect asymptotic properties. 
We fix this problem in two alternative ways. First, by a
Woods-Saxon (WS) potential solution fit to the interior of the NCSM
overlap integrals.
Second, by a direct matching with the Whittaker function.
The corrected overlap integrals are then used for the
$^7$Be(p,$\gamma$)$^8$B S-factor calculation. We study the
convergence of the S-factor with respect to the NCSM HO frequency
and the model space size. Our S-factor results are in agreement 
with recent direct measurement data. 
We also test the spectroscopic factors and the corrected
overlap integrals from the NCSM in describing the momentum
distributions in knockout reactions with $^8$B projectiles. A good
agreement with the available experimental data is also found,
attesting the overall consistency of the calculations.
\end{abstract}
\maketitle
%
\section{\label{sec:intro}Introduction}
The $^7$Be(p,$\gamma$)$^8$B capture reaction serves as an important input
for understanding the solar neutrino flux \cite{Adelberger}.
Recent experiments determined the neutrino flux emitted from $^8$B
with a precision of ~9\% \cite{SNO}. On the other hand, theoretical predictions
have uncertainties of the order of 20\% \cite{CTK03,BP04}.
The theoretical neutrino flux depends on the $^7$Be(p,$\gamma$)$^8$B S-factor
that needs to be known with high precision.
Many experimental and theoretical investigations were devoted to this reaction.
Experiments were performed using direct measurement techniques with proton beam
and $^7$Be targets \cite{Filippone,Baby,Seattle} as well as by indirect methods
when a $^8$B beam breaks up into $^7$Be and proton in the Coulomb field of
a heavy target \cite{BBR86,Be7pgamm_exp}.
Theoretical calculations needed to extrapolate the measured S-factor to the astrophysically
relevant Gamow energy were performed with several methods: the R-matrix parametrization
\cite{Barker95}, the potential model \cite{Robertson,Typel97,Davids03}, and the microscopic
cluster models \cite{DB94,Csoto95,D04}.

In this work, we present the first calculation of the
$^7$Be(p,$\gamma$)$^8$B S-factor starting from the {\it ab initio}
wave functions of $^8$B and $^7$Be. We apply the {\it ab initio}
no-core shell model (NCSM) \cite{NCSMC12,NKB00}. In this method, one considers nucleons
interacting by high-precision nucleon-nucleon (NN) potentials. There
are no adjustable or fitted parameters. Within the NCSM, we study
the binding energies and other nuclear structure properties of
$^7$Be, $^8$B as well as $^{7,8}$Li, and calculate overlap integrals
for the $^8$B and $^7$Be bound states. Due to the use of the
harmonic-oscillator (HO) basis, we have to correct the asymptotic
behavior of the NCSM overlap integrals. This is done in two alternative ways. 
First, by fitting
Woods-Saxon (WS) potential solutions to the interior part of the NCSM
overlap integrals under the constraint that the experimental
$^7$Be+p threshold is reproduced. 
Second, by a direct matching of the NCSM overlap integrals
and the Whittaker function.
The corrected overlap integrals
are then utilized to calculate the $^7$Be(p,$\gamma$)$^8$B S-factor
as well as momentum distributions in stripping reactions. We pay
special attention to the convergence of the S-factor with respect to
the NCSM model space size and the HO frequency.

In Sect.~\ref{sec:spectroscopy}, we present the NCSM results for
$^7$Be, $^8$B and $^{7,8}$Li energies, ground-state radii and
electromagnetic moments and transitions. The calculation of cluster
form factors and the correction of their asymptotics is described in
Sect.~\ref{sec:overlap}. A test of the corrected NCSM overlaps and
spectroscopic factors for the momentum distributions in the
stripping reaction $(^7{\rm Be}+p)+A\longrightarrow \ ^7{\rm Be}+X$
is discussed in Sect.~\ref{sec:momdis}. The S-factor calculation
with sensitivity and convergence studies is presented in
Sect.~\ref{sec:sfactor}. Conclusions are drawn in
Sect.~\ref{sec:conc}.

\section{\label{sec:spectroscopy}Nuclear structure of $^7$Be, $^7$Li, $^8$B and $^8$Li}
\subsection{\label{subsec:NCSM} {\it Ab initio} no-core shell model}

In the NCSM, we consider a system of $A$ point-like non-relativistic
nucleons that interact by realistic two- or two- plus three-nucleon
interactions. The calculations are performed using a finite harmonic
oscillator (HO) basis. As in the present application we aim at
describing loosely bound states, it is desirable to include as many
terms as possible in the expansion of the total wave function. By
restricting our study to two-nucleon (NN) interactions, even though
the NCSM allows for the inclusion of three-body forces \cite{v3b},
we are able to maximize the model space and to better observe the
convergence of our results. The NCSM theory was outlined in many
papers. Here, we only briefly review the main points for the case
when the Hamiltonian is restricted to just a two-nucleon
interaction.

We start from the intrinsic two-body Hamiltonian for the $A$-nucleon system
$H_A=T_{rel} + {\cal V}$, where
$T_{rel}$ is the relative kinetic energy
and ${\cal V}$ is the sum of two-body nuclear and Coulomb
interactions. Since we solve the many-body problem
in a finite HO basis space,
it is necessary that we derive a model-space dependent effective
Hamiltonian. For this purpose, we perform
a unitary transformation \cite{NCSMC12,NKB00,LS81,UMOA}
of the Hamiltonian, which accommodates the short-range correlations.
In general,
the transformed Hamiltonian is an $A$-body operator.
Our simplest, yet non-trivial, approximation
is to develop a two-particle cluster effective Hamiltonian, while
the next improvement is to include three-particle clusters, and so on.
The effective interaction is then obtained
from the decoupling condition between the model space and the excluded space
for the two-nucleon transformed Hamiltonian.
Details of the procedure are
described in Refs.~\cite{NCSMC12,NKB00}.
The resulting two-body effective Hamiltonian
depends on the nucleon number $A$, the HO frequency $\Omega$, and
$N_{\rm max}$, the maximum many-body HO excitation energy
defining the model space.
It follows that the effective interaction approaches the starting bare
interaction for $N_{\rm max}\rightarrow \infty$.
Our effective interaction is translationally invariant.
A significant consequence of this fact is
the factorization of our wave functions into a product of a center-of-mass
$\frac{3}{2}\hbar\Omega$
component times an internal component.

Our most significant approximation here
is the use of the two-body cluster approximation to
the  effective many-body Hamiltonian.
Our method is not variational so higher-order terms
may contribute with either sign to total binding.
Hence, evaluating the dependence on the basis-space parameters
help calibrate our convergence.

Once the effective interaction
is derived, we diagonalize the effective Hamiltonian
in a Slater determinant HO basis that spans a complete
$N_{\rm max}\hbar\Omega$ space.
This is a highly non-trivial problem due to very large dimensions
we encounter. To solve this problem we have
used a specialized version of the shell model code
Antoine~\cite{EC99,Antoine}, recently adapted to the
NCSM~\cite{Be8,A10_NCSM}. This code works in the $M$ scheme for basis
states, and uses the Lanczos algorithm for diagonalization. 
Its basic idea is to write the basis
states as a product of two Slater determinants, a proton one 
and a neutron one.
Matrix elements of operators are calculated for each separate subspace
(one-body for the proton-neutron, two-body for the proton-proton 
and neutron-neutron).
The performance of the code is the best when the ratio between the number
of proton plus neutron Slater determinants and the dimension of the matrix is the least. 
It happens when the number of proton Slater determinants 
is equal the number of the neutron Slater determinants. The number
of iterations needed to converge the first eigenstates is significantly
reduced by the implementation of a sophisticated strategy for selecting
the pivot vectors.
The difficulty of the no-core
calculations (in which all nucleons are active) is that the number of shells
and, consequently, the number of matrix elements that are precalculated,
becomes very large. One has to handle a huge number of operators. 
This is the reason why it has been necessary to write a specialized version 
of the code.

A recent development of the NCSM is the ability to further process the
wave functions, resulting from the shell-model calculation, to obtain
channel cluster form factors~\cite{cluster} and, consequently, the spectroscopic
factors.

\subsection{\label{subsec:NNpot}Nucleon-nucleon interactions}

Two different, high-precision (i.e. such that provide a perfect fit to two-nucloen data)
NN interactions have been used in this
study: the CD-Bonn 2000 ~\cite{cdb2k} and the INOY
(Inside Nonlocal Outside Yukawa)~\cite{dol03:67,dol04:69} potentials.

The CD-Bonn 2000 potential \cite{cdb2k} as well as its earlier version \cite{cdb}
is a charge-dependent NN interaction based
on one-boson exchange. It is described in terms of covariant Feynman
amplitudes, which are non-local. Consequently, the off-shell behavior of
the CD-Bonn interaction differs from local potentials
which leads to larger binding energies in nuclear few-body systems.

A new type of interaction, which respects the local behavior of
traditional NN interactions at longer ranges but exhibits a
non-locality at shorter distances, was recently proposed by
Doleschall \emph{et al.}~\cite{dol03:67,dol04:69}. The authors
explore the extent to which effects of multi-nucleon forces can be
absorbed by non-local terms in the NN interaction. Their goal was to
investigate if it is possible to introduce non-locality in the NN
interaction so that it correctly describes the three-nucleon bound
states $^{3}$H and $^{3}$He, while still reproducing NN scattering
data with high precision. The so called IS version of this
interaction, introduced in Ref.~\cite{dol03:67}, contains
short-range non-local potentials in $^1S_0$ and $^3S_1-^{3\!\!}D_1$
partial waves while higher partial waves are taken from Argonne
$v_{18}$. In this study we are using the IS-M version, which
includes non-local potentials also in the $P$ and $D$
waves~\cite{dol04:69}. It is important to note that, for this
particular version, the on-shell properties of the triplet $P$-wave
interactions have been modified in order to improve the description
of $3N$ analyzing powers. The $^{3\!}P_0$ interaction was adjusted
to become less attractive, the $^3P_1$ became more repulsive, and
the $^{3\!}P_2$ more attractive. Unfortunately, this gives a
slightly worse fit to the Nijmegen $^{3\!}P$ phase shifts.

\subsection{\label{subsec:A_7}$^7$Be and $^7$Li}

Our calculations for both $A=7$ and $A=8$ nuclei were performed in model spaces
up to $10\hbar\Omega$ for a wide range of HO frequencies. We then selected the optimal HO frequency
corresponding to the ground-state energy minimum in the $10\hbar\Omega$ space and performed
a $12\hbar\Omega$ calculation to obtain the ground-state energy and the point-nucleon 
root-mean-square (rms) radii.
The overlap integrals as well as other observables were, however, calculated only using wave
functions from up to $10\hbar\Omega$ spaces.

\begin{figure}[hbtp]
  \includegraphics*[width=0.55\columnwidth,angle=90]
   {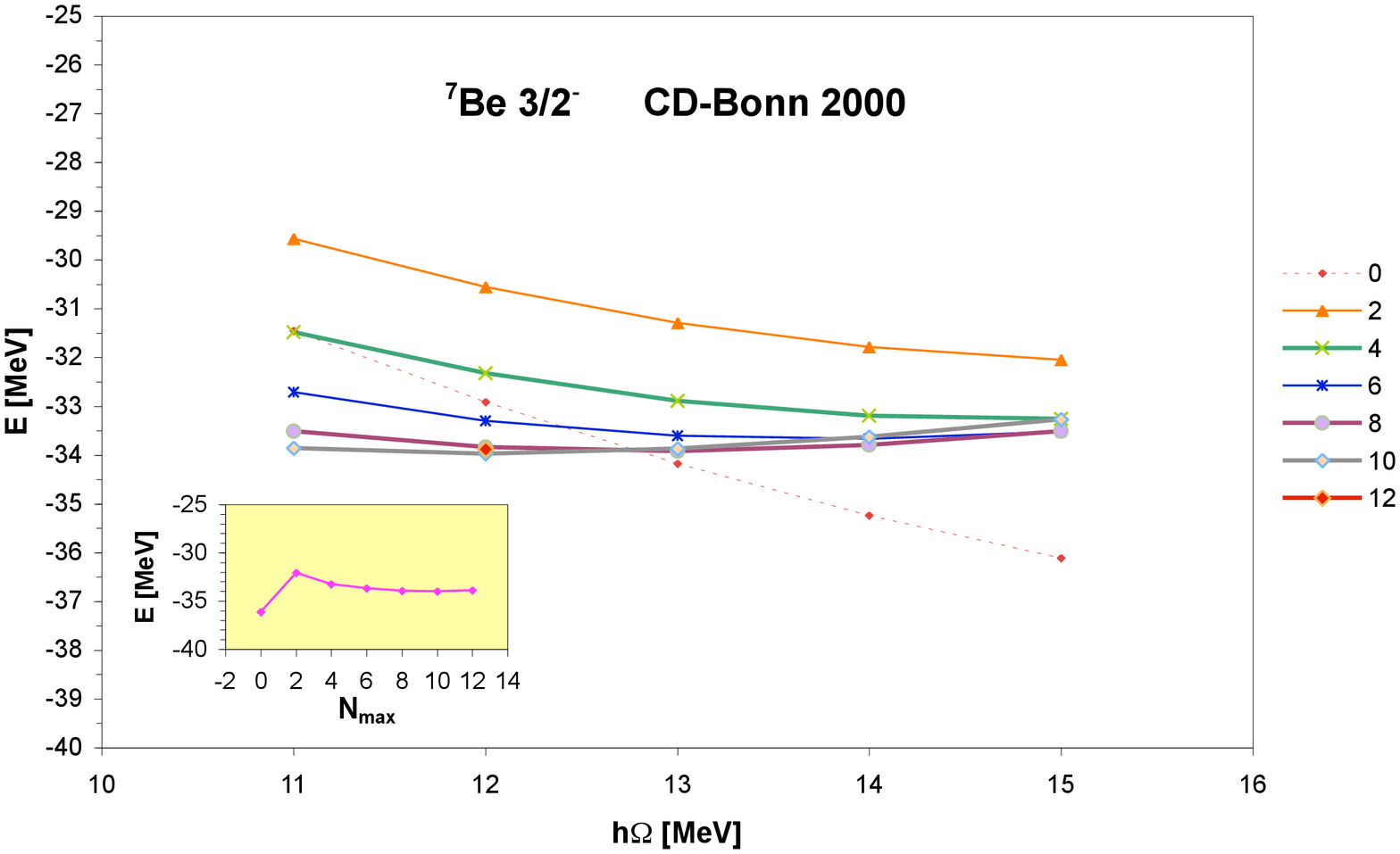}
  \caption{HO frequency dependence of the $^7$Be ground-state energy
for model spaces from $0\hbar\Omega$ to $12\hbar\Omega$ obtained using
the CD-Bonn 2000 NN potential. The inset demonstrates how the values
at the minima of each curve converge with increasing $N_{\rm max}$.
  \label{Be7_cdb2k_gs}}
\end{figure}

\begin{figure}[hbtp]
  \includegraphics*[width=0.55\columnwidth,angle=90]
   {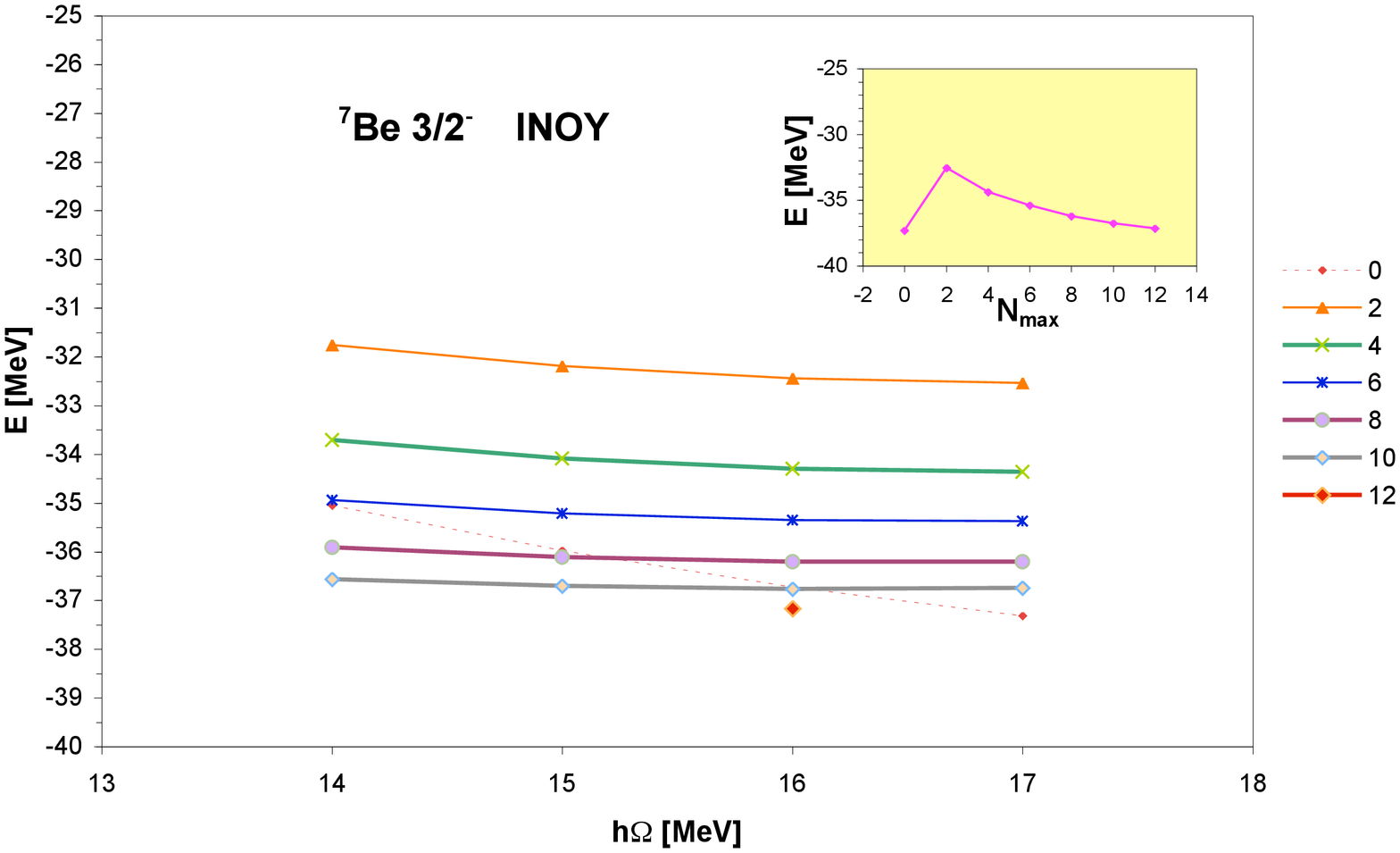}
  \caption{Same as in Fig.\ref{Be7_cdb2k_gs}, but using the INOY NN potential.
  \label{Be7_INOY_gs}}
\end{figure}

The $^7$Be ground-state energy dependence on the HO frequency for different model spaces
is shown in Figs.~\ref{Be7_cdb2k_gs} and \ref{Be7_INOY_gs} for the CD-Bonn 2000 and the INOY NN
potentials, respectively. We observe a quite different convergence trend for the two potentials.
For the INOY, the convergence is very uniform with respect to the HO frequency with substantial
changes with $N_{\rm max}$. The convergence with increasing $N_{\rm max}$ is evident as also seen in the
inset of the Fig. \ref{Be7_INOY_gs}. It is straightforward to extrapolate that the converged
INOY ground-state energy will slightly overbind $^7$Be. The ground-state energy
convergence for the CD-Bonn
is quite different with a stronger dependence on the frequency, with minima shifting to smaller
frequency with basis size increase, and an overall weaker dependence on $N_{\rm max}$.
Contrary to the INOY, the CD-Bonn underbinds $^7$Be by more than 3 MeV, which is typical for the
standard high-precision NN potentials.


\begin{figure}[hbtp]
  \includegraphics*[width=0.9\columnwidth]
   {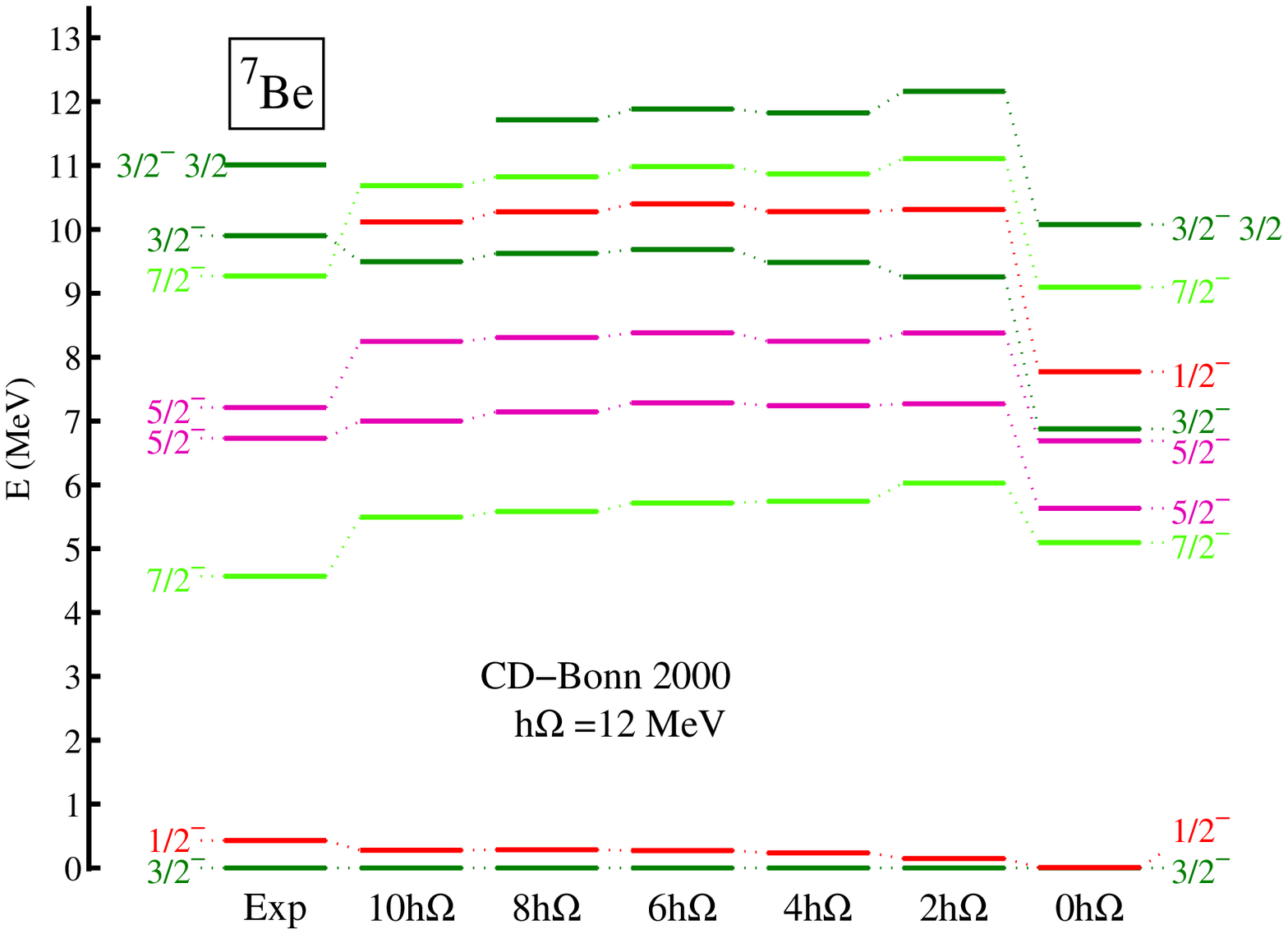}
  \caption{Basis size dependence of the $^7$Be excitation spectrum in the range from
$0\hbar\Omega$ to $10\hbar\Omega$ obtained using the CD-Bonn 2000 NN potential
and the HO frequency of $\hbar\Omega=12$ MeV.%
  \label{Be7_cdb2k_12}}
\end{figure}

\begin{figure}[hbtp]
  \includegraphics*[width=0.9\columnwidth]
   {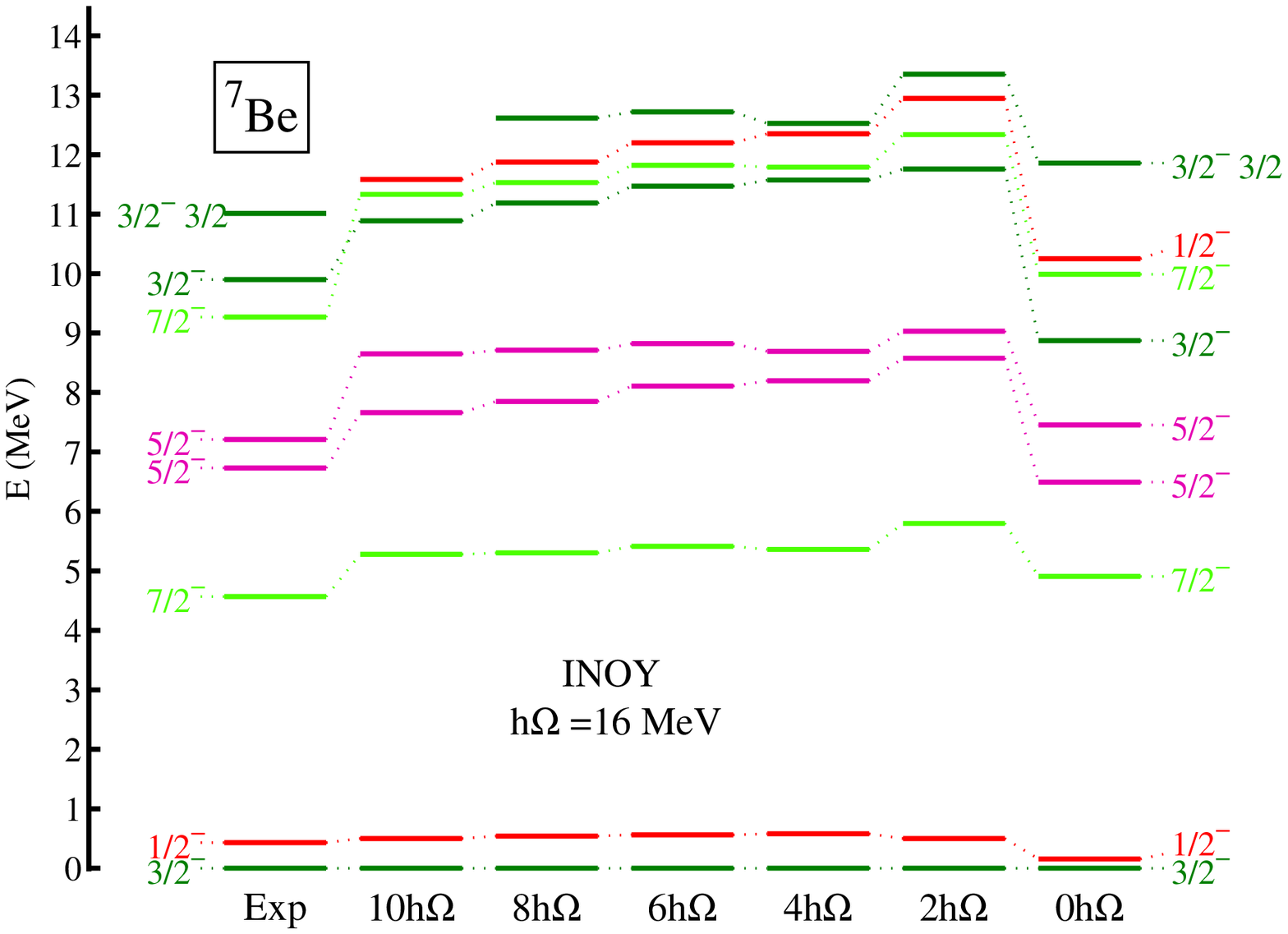}
  \caption{Same as in Fig.~\ref{Be7_cdb2k_12}, but using the INOY NN potential
and the HO frequency of $\hbar\Omega=16$ MeV.%
  \label{Be7_INOY_16}}
\end{figure}

From these results we select the optimal frequency $\hbar\Omega=12$ MeV
for the CD-Bonn 2000 and $\hbar\Omega=16$ MeV for the INOY
potential. Corresponding spectra obtained using these frequencies
are then shown in Figs.~\ref{Be7_cdb2k_12} and \ref{Be7_INOY_16}.
The energies, radii and electromagnetic observables are summarized
in Tables~\ref{tab:Be7energies} and \ref{tab:Be7moments}, where we
also include the $^7$Li results. We obtain the same level ordering
for $^7$Be and $^7$Li which is also the same for both NN potentials
with the exception of a reversal of the $7/2^-_2$ and $1/2^-_2$
levels. Our CD-Bonn ordering is in agreement with experiment for the
9 lowest levels in $^7$Li. In $^7$Be, the experimental $7/2^-_2$ and
$3/2^-_2$ levels are reversed compared to our results and the
situation in $^7$Li. Our calculated spectra show a good convergence
with the basis size enlargement as well as a good stability with
respect to the frequency change. An interesting feature is a
different level spacing of the $7/2^-_1$, $5/2^-_1$ and $5/2^-_2$
levels for the CD-Bonn and INOY calculations. The INOY results are
similar to those obtained when a three-body interaction is included
in the Hamiltonian \cite{v3b}. This is not surprising as discussed
in Subsect. \ref{subsec:NNpot}.

\begin{table}[hbtp]
  \caption{The $^7$Be and $^7$Li ground and excited state energies (in MeV)
obtained using the CD-Bonn 2000 and INOY NN potentials.
The HO frequency of $\hbar\Omega=12$(16) MeV was used
in the CD-Bonn 2000 (INOY) NN potential calculation.
The ground (excited) state energies were
obtained in the $12\hbar\Omega$ ($10\hbar\Omega$) model space.
Experimental values are from Refs. \cite{A=5-7}.
  \label{tab:Be7energies}}
  \begin{ruledtabular}
    \begin{tabular}{cccc}
\multicolumn{4}{c}{$^7$Be} \\
                                          & Expt.   & CD-Bonn 2000 & INOY \\
$|E_{\rm gs}(\frac{3}{2}^- \frac{1}{2})|$ &  37.6004(5) & 33.881 & 37.161 \\
$E_{\rm x}(\frac{3}{2}^-_1 \frac{1}{2})$ & 0.0      & 0.0    & 0.0   \\
$E_{\rm x}(\frac{1}{2}^-_1 \frac{1}{2})$ & 0.429    & 0.278  & 0.501 \\
$E_{\rm x}(\frac{7}{2}^-_1 \frac{1}{2})$ & 4.57(5)  & 5.494  & 5.278 \\
$E_{\rm x}(\frac{5}{2}^-_1 \frac{1}{2})$ & 6.73(10) & 6.999  & 7.660 \\
$E_{\rm x}(\frac{5}{2}^-_2 \frac{1}{2})$ & 7.21(6)  & 8.247  & 8.648 \\
$E_{\rm x}(\frac{7}{2}^-_2 \frac{1}{2})$ & 9.27(10) & 10.687 & 11.331 \\
$E_{\rm x}(\frac{3}{2}^-_2 \frac{1}{2})$ & 9.9      & 9.493  & 10.887 \\
$E_{\rm x}(\frac{1}{2}^-_2 \frac{1}{2})$ &          & 10.120 & 11.583 \\
$E_{\rm x}(\frac{3}{2}^-_1 \frac{3}{2})$ & 11.01(3) & 11.717 & 12.607 \\
\hline
\multicolumn{4}{c}{$^7$Li} \\
                                          & Expt.   & CD-Bonn 2000 & INOY \\
$|E_{\rm gs}(\frac{3}{2}^- \frac{1}{2})|$ & 39.245  & 35.524 & 38.892 \\
$E_{\rm x}(\frac{3}{2}^-_1 \frac{1}{2})$ & 0.0      & 0.0    & 0.0   \\
$E_{\rm x}(\frac{1}{2}^-_1 \frac{1}{2})$ & 0.478    & 0.285  & 0.513 \\
$E_{\rm x}(\frac{7}{2}^-_1 \frac{1}{2})$ & 4.65     & 5.585  & 5.353 \\
$E_{\rm x}(\frac{5}{2}^-_1 \frac{1}{2})$ & 6.60     & 7.079  & 7.741 \\
$E_{\rm x}(\frac{5}{2}^-_2 \frac{1}{2})$ & 7.45     & 8.522  & 8.902 \\
$E_{\rm x}(\frac{3}{2}^-_2 \frac{1}{2})$ & 8.75     & 9.849  & 11.267 \\
$E_{\rm x}(\frac{1}{2}^-_2 \frac{1}{2})$ & 9.09     & 10.458 & 11.931 \\
$E_{\rm x}(\frac{7}{2}^-_2 \frac{1}{2})$ & 9.57     & 11.033 & 11.691 \\
$E_{\rm x}(\frac{3}{2}^-_1 \frac{3}{2})$ & 11.24    & 11.974 & 12.826 \\
    \end{tabular}
  \end{ruledtabular}
\end{table}

Concerning the magnetic moments
and M1 transitions, we can see a very little dependence of the calculated values
on the HO frequency or the basis size. Also the two NN potentials give very similar results that are
in a reasonable agreement with experiment. Concerning the radii and quadrupole moments,
the calculated values in general increase with increasing basis size and decreasing frequency.
This is in part a consequence of the incorrect asymptotics of the HO basis. The fastest convergence
for the radii and quadrupole moment occurs at a smaller HO frequency. In the CD-Bonn calculations
for $\hbar\Omega=11$ and $12$ MeV, the radii are close to experimental values. The quadrupole moments
are still underestimated. For the INOY potential we observe underestimation of both radii
and quadrupole moments. This is not unexpected as the INOY NN potential also underpredicts the 
$^4$He radius.

\begin{table}[hbtp]
  \caption{The $^7$Be and $^7$Li point-proton rms radii (in fm), ground-state quadrupole (in $e$fm$^2$)
and magnetic (in $\mu_{\rm N}$) moments and M1 transitions (in $\mu_{\rm N}^2$)
obtained within the NCSM for different HO frequencies (given in MeV)
and model spaces for the CD-Bonn 2000 and INOY NN potentials. Experimental values
are from Refs. \cite{A=5-7}.
  \label{tab:Be7moments}}
  \begin{ruledtabular}
    \begin{tabular}{cccccc}
\multicolumn{6}{c}{$^7$Be} \\
\multicolumn{6}{c}{CD-Bonn 2000} \\
$\hbar\Omega$ & $N_{\rm max}$ & $r_{\rm p}$ & Q & $\mu$ & B(M1; $\frac{1}{2}^-\rightarrow
\frac{3}{2}^-$)\\
\hline
12 &  6 & 2.311 & -4.755 & -1.150 & 3.192 \\
12 &  8 & 2.324 & -4.975 & -1.151 & 3.145 \\
12 & 10 & 2.342 & -5.153 & -1.141 & 3.114 \\
12 & 12 & 2.365 &  &  &\\
11 &  6 & 2.377 &  -5.029 & -1.150 & 3.203 \\
11 &  8 & 2.377 &  -5.186 & -1.151 & 3.162 \\
11 & 10 & 2.383 &  -5.354 & -1.155 & 3.125 \\
\multicolumn{2}{c}{Expt.} & 2.36(2) &   & -1.398(15) & 3.71(48) \\
\multicolumn{6}{c}{INOY} \\
$\hbar\Omega$ & $N_{\rm max}$ & $r_{\rm p}$ & Q & $\mu$ & B(M1; $\frac{1}{2}^-\rightarrow
\frac{3}{2}^-$)\\
\hline
16 &  6 & 2.114 & -3.946 & -1.161 & 3.158 \\
16 &  8 & 2.149 & -4.212 & -1.157 & 3.119 \\
16 & 10 & 2.181 & -4.459 & -1.151 & 3.092 \\
16 & 12 & 2.214 &  &  &\\
14 &  6 & 2.190 & -4.245 & -1.162 & 3.180 \\
14 &  8 & 2.209 & -4.454 & -1.158 & 3.139 \\
14 & 10 & 2.227 & -4.654 & -1.155 & 3.108 \\
\multicolumn{2}{c}{Expt.} & 2.36(2) &   & -1.398(15) & 3.71(48) \\
\hline
\multicolumn{6}{c}{$^7$Li} \\
\multicolumn{6}{c}{CD-Bonn 2000} \\
$\hbar\Omega$ & $N_{\rm max}$ & $r_{\rm p}$ & Q & $\mu$ & B(M1; $\frac{1}{2}^-\rightarrow
\frac{3}{2}^-$)\\
\hline
12 &  6 & 2.149 & -2.717 & +3.027 & 4.256 \\
12 &  8 & 2.156 & -2.866 & +3.020 & 4.188 \\
12 & 10 & 2.168 & -3.001 & +3.011 & 4.132 \\
12 & 12 & 2.188 & -3.130 & & \\
\multicolumn{2}{c}{Expt.} & 2.27(2) & -4.06(8) & +3.256 & 4.92(25) \\
\multicolumn{6}{c}{INOY} \\
$\hbar\Omega$ & $N_{\rm max}$ & $r_{\rm p}$ & Q & $\mu$ & B(M1; $\frac{1}{2}^-\rightarrow
\frac{3}{2}^-$)\\
\hline
16 &  6 & 1.963 & -2.334 & +3.039 & 4.192 \\
16 &  8 & 1.990 & -2.500 & +3.029 & 4.136 \\
16 & 10 & 2.015 & -2.648 & +3.021 & 4.098 \\
16 & 12 & 2.042 & -2.788 & & \\
\multicolumn{2}{c}{Expt.} & 2.27(2) & -4.06(8) & +3.256 & 4.92(25) \\
    \end{tabular}
  \end{ruledtabular}
\end{table}

\subsection{\label{subsec:A_8}$^8$B and $^8$Li}

\begin{figure}[hbtp]
  \includegraphics*[width=0.55\columnwidth,angle=90]
   {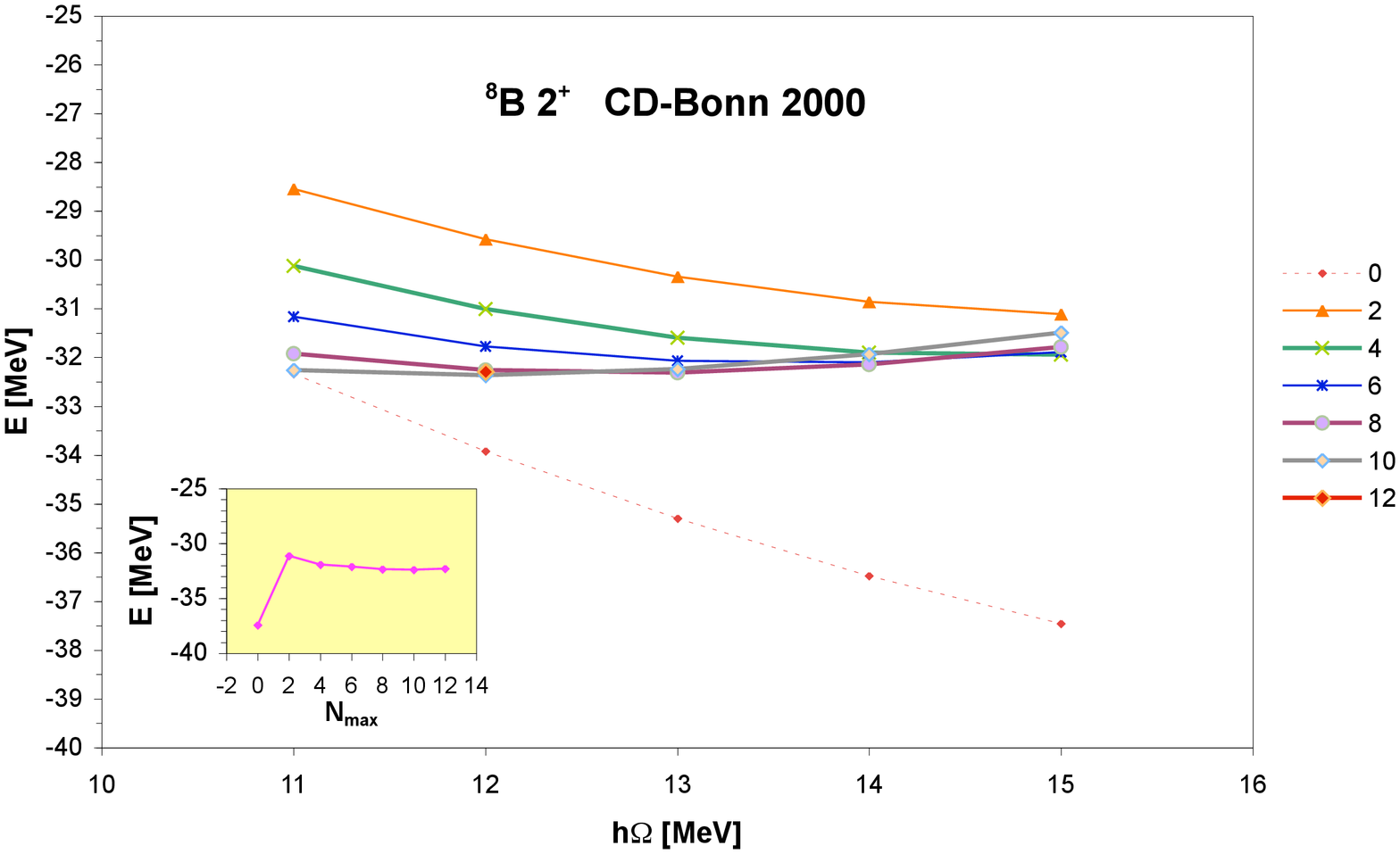}
  \caption{HO frequency dependence of the $^8$B ground-state energy
for model spaces from $0\hbar\Omega$ to $12\hbar\Omega$ obtained using
the CD-Bonn 2000 NN potential. The inset demonstrates how the values
at the minima of each curve converge with increasing $N_{\rm max}$.
  \label{B8_cdb2k_gs}}
\end{figure}

\begin{figure}[hbtp]
  \includegraphics*[width=0.55\columnwidth,angle=90]
   {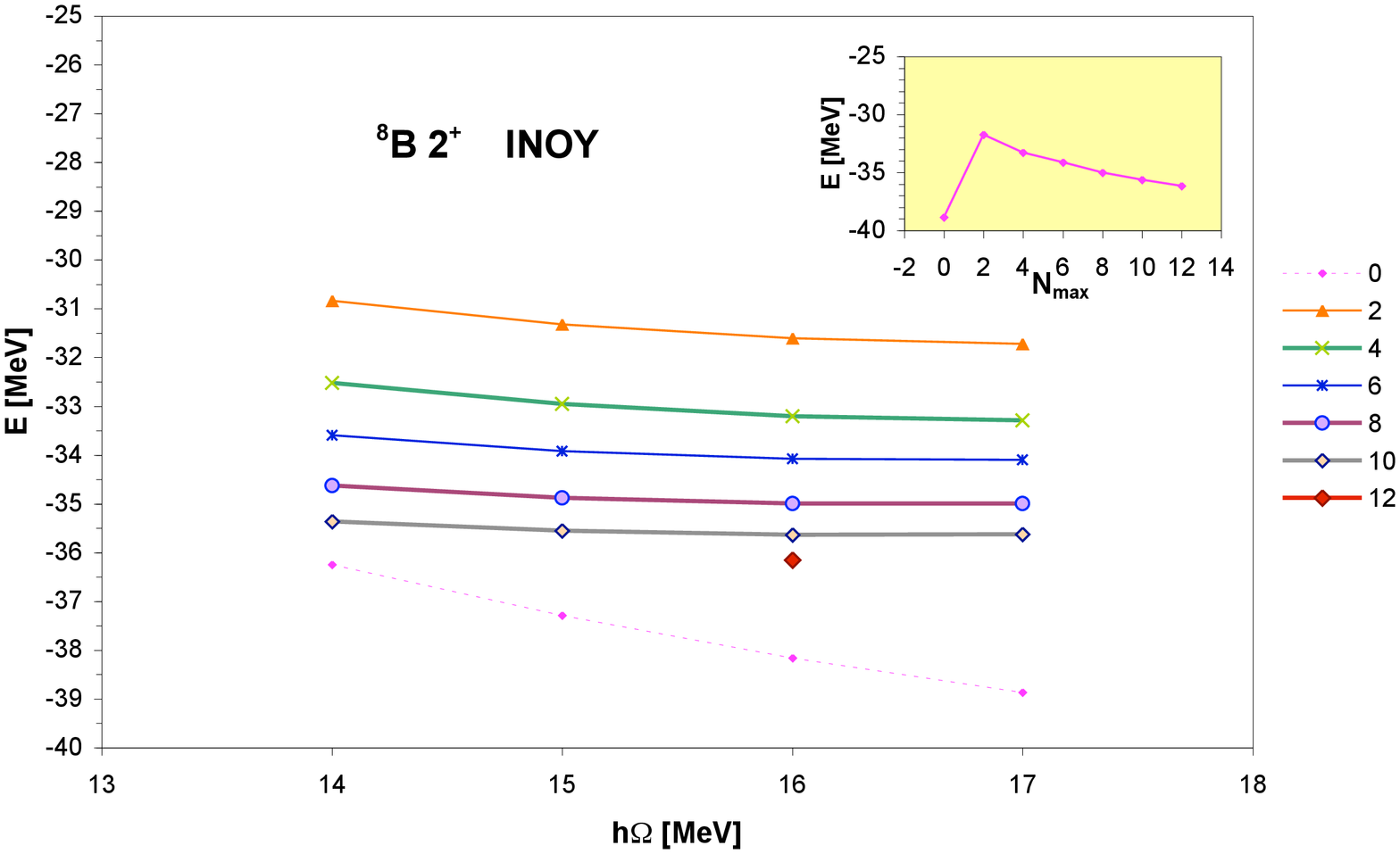}
  \caption{Same as in Fig.~\ref{B8_cdb2k_gs}, but using the INOY NN potential.
  \label{B8_INOY_gs}}
\end{figure}

\begin{figure}[hbtp]
  \includegraphics*[width=0.9\columnwidth]
   {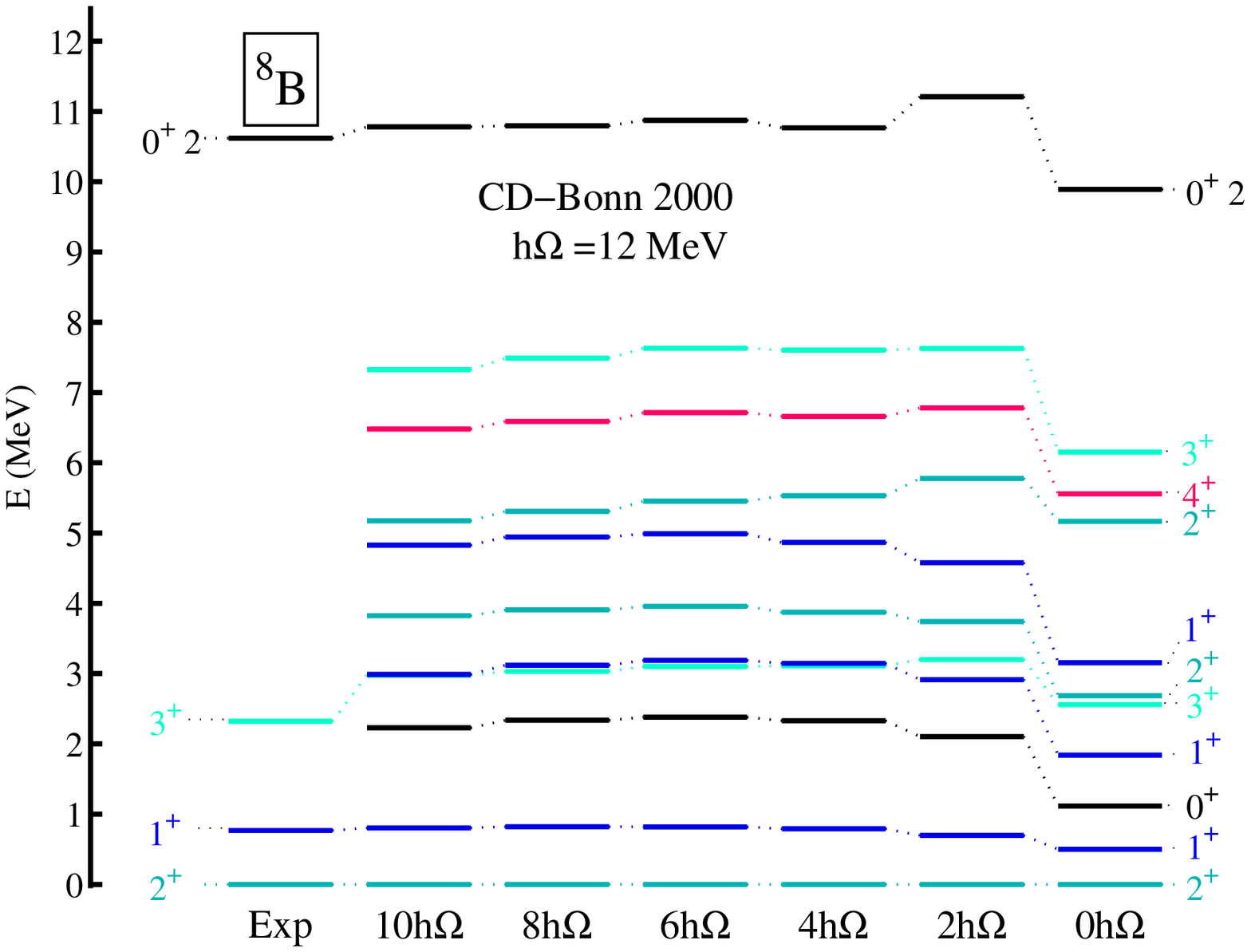}
  \caption{Basis size dependence of the $^8$B excitation spectrum in the range from
$0\hbar\Omega$ to $10\hbar\Omega$ obtained using the CD-Bonn 2000 NN potential
and the HO frequency of $\hbar\Omega=12$ MeV.%
  \label{B8_cdb2k_12}}
\end{figure}

\begin{figure}[hbtp]
  \includegraphics*[width=0.9\columnwidth]
   {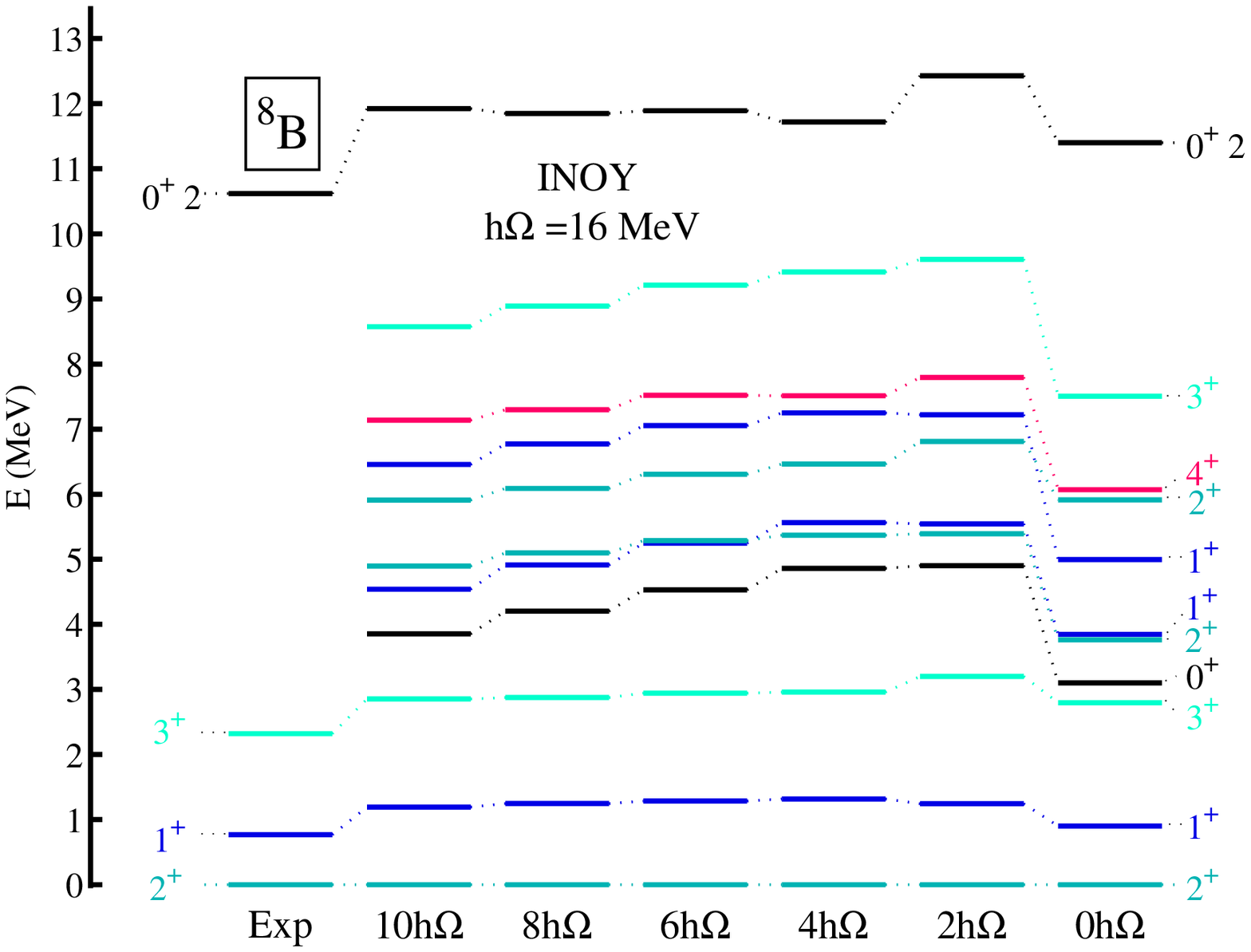}
  \caption{Same as in Fig.~\ref{B8_cdb2k_12}, but using the INOY NN potential
and the HO frequency of $\hbar\Omega=16$ MeV.%
  \label{B8_INOY_16}}
\end{figure}

Our $^8$B ground-state energy results are presented in Figs. \ref{B8_cdb2k_gs}
and \ref{B8_INOY_gs} as well as in Table \ref{tab:B8energies} where
we also show the excitation energies and the corresponding $^8$Li results.
The basis size dependence of the $^8$B spectra calculated using the optimal
HO frequencies is shown in Figs. \ref{B8_cdb2k_12}
and \ref{B8_INOY_16} for the CD-Bonn and INOY NN potentials, respectively.
Similar conclusions can be drawn as for the $A=7$ nuclei concerning convergence.
The INOY NN potential gives binding energy close to the experimental one
and a small overbinding is expected based on extrapolation of our $N_{\rm max}$
dependence. The CD-Bonn underbinds both $^8$B and $^8$Li by about 5 MeV.
We note that the $A=8$ nuclei, with emphasis on $^8$Be, were also investigated
within the NCSM in Ref.~\cite{Be8}. The present $^8$B and $^8$Li CD-Bonn binding energy
results are basically identical to those of Ref.~\cite{Be8} with small differences
due to the use of different versions, \cite{cdb2k} vs. \cite{cdb}, of the CD-Bonn NN
potential.
It is interesting to note that both employed NN potentials predict $^8$B unbound,
contrary to experiment. The problem is less severe for the INOY
NN potential. This suggests that a three-nucleon interaction is essential to
reproduce the experimental threshold. This may appear as a problem in view
of the present application to $^7$Be(p,$\gamma$)$^8$B reaction. However, since
our basis has incorrect asymptotics in the first place, we make use of only the interior
part of our {\it ab initio} wave functions that are presumably unaffected by
the wrong threshold placements. This is discussed in the next section.


Concerning the ground-state energy systematics of the $A=7$ and $A=8$ nuclei, 
we note that recent Green's function Monte Carlo (GFMC) calculations for $^7$Li and $^8$Li 
using the Argonne potentials \cite{GFMC_04} found qualitatively similar differences 
between the calculated and experimental values as we observe here.

\begin{table}[hbtp]
  \caption{The $^8$B and $^8$Li ground and excited state energies (in MeV)
obtained using the CD-Bonn 2000 and INOY NN potentials.
The HO frequency of $\hbar\Omega=12$(16) MeV was used
in the CD-Bonn 2000 (INOY) NN potential calculation.
The ground (excited) state energies were
obtained in the $12\hbar\Omega$ ($10\hbar\Omega$) model space.
Experimental values are from Refs. \cite{A=8}.
  \label{tab:B8energies}}
  \begin{ruledtabular}
    \begin{tabular}{cccc}
\multicolumn{4}{c}{$^8$B} \\
                         & Expt.   & CD-Bonn 2000 & INOY \\
$|E_{\rm gs}(2^+ 1)|$    & 37.7378(11) & 32.284 & 36.148 \\
$E_{\rm x}(2^+_1 1)$       &  0.0      & 0.0    & 0.0   \\
$E_{\rm x}(1^+_1 1)$       &  0.774(6) & 0.804  & 1.199 \\
$E_{\rm x}(3^+_1 1)$       &  2.32(20) & 2.977  & 2.854 \\
$E_{\rm x}(0^+_1 1)$       &           & 2.229  & 3.853 \\
$E_{\rm x}(1^+_2 1)$       &           & 2.988  & 4.540 \\
$E_{\rm x}(2^+_2 1)$       &           & 3.824  & 4.897 \\
$E_{\rm x}(1^+_3 1)$       &           & 4.827  & 6.459 \\
$E_{\rm x}(2^+_3 1)$       &           & 5.175  & 5.908 \\
$E_{\rm x}(4^+_1 1)$       &           & 6.482  & 7.138 \\
$E_{\rm x}(3^+_2 1)$       &           & 7.325  & 8.572 \\
$E_{\rm x}(0^+_1 2)$       & 10.619(9) & 10.782 & 11.926 \\
\hline
\multicolumn{4}{c}{$^8$Li} \\
                           & Expt.   & CD-Bonn 2000 & INOY \\
$|E_{\rm gs}(2^+ 1)|$      &  41.277 & 35.820   & 39.938 \\
$E_{\rm x}(2^+_1 1)$       &  0.0      & 0.0    & 0.0   \\
$E_{\rm x}(1^+_1 1)$       &  0.981    & 0.855  & 1.264 \\
$E_{\rm x}(3^+_1 1)$       &  2.255(3) & 3.019  & 2.871 \\
$E_{\rm x}(0^+_1 1)$       &           & 2.480  & 4.225 \\
$E_{\rm x}(1^+_2 1)$       &  3.21     & 3.247  & 4.903 \\
$E_{\rm x}(2^+_2 1)$       &           & 3.977  & 5.114 \\
$E_{\rm x}(1^+_3 1)$       &           & 5.023  & 6.758 \\
$E_{\rm x}(2^+_3 1)$       &           & 5.290  & 6.071 \\
$E_{\rm x}(4^+_1 1)$       &  6.53(20) & 6.691  & 7.398 \\
$E_{\rm x}(3^+_2 1)$       &           & 7.570  & 8.915 \\
$E_{\rm x}(0^+_1 2)$       &  10.822   & 10.898 & 12.049 \\
    \end{tabular}
  \end{ruledtabular}
\end{table}

Concerning the excitation energies, a noticeable difference between
the CD-Bonn and the INOY predictions for the low-lying levels is a
reversed order of the $0^+_1$ and $3^+_1$ states. We note, however,
that the $0^+_1$ state has not been observed experimentally.


As seen from Table \ref{tab:B8moments}, the radii and quadrupole
moments are substantially larger and closer to experiment in our
CD-Bonn calculations. Contrary to the $^7$Be-$^7$Li case, here we
observe an interesting difference between the two NN potentials for
the magnetic moment prediction. The INOY NN potential gives the
magnetic moment of $^8$Li greater than that of $^8$B in agreement
with experiment, while the CD-Bonn predicts the opposite.
Interestingly, almost identical $^8$Li and $^8$B magnetic moments as
we obtained using the CD-Bonn were reported in
Ref.~\cite{D04} calculated within a cluster model applied
to the $^7$Be(p,$\gamma$)$^8$B reaction. Clearly, our results
suggest that the $A=8$ magnetic moments are sensitive to a presence
of three-nucleon interaction in the Hamiltonian (that is mocked up
to certain degree by the INOY NN potential). As to the
B(M1;$1^+\rightarrow 2^+$) transition, both potentials gives about
the same result that is in agreement with the $^8$Li experimental
value but by almost a factor of three smaller than the experimental
value for $^8$B. We note the large experimental error bar of the
latter. It appears that the $^8$B experimental values is
inconsistent with both the analogous $^8$Li value and our
calculations.

\begin{table}[hbtp]
  \caption{The $^8$B and $^8$Li point-proton rms radii (in fm), ground-state quadrupole (in $e$fm$^2$)
and magnetic (in $\mu_{\rm N}$) moments and M1 transitions (in $\mu_{\rm N}^2$)
obtained within the NCSM for different HO frequencies (given in MeV)
and model spaces for the CD-Bonn 2000 and INOY NN potentials. Experimental values
are from Refs. \cite{A=8}.
  \label{tab:B8moments}}
  \begin{ruledtabular}
    \begin{tabular}{cccccc}
\multicolumn{6}{c}{$^8$B} \\
\multicolumn{6}{c}{CD-Bonn 2000} \\
$\hbar\Omega$ & $N_{\rm max}$ & $r_{\rm p}$ & Q & $\mu$ & B(M1; $1^+\rightarrow 2^+$) \\
\hline
12 &  6 & 2.436 & +5.218 & +1.463 & 3.498 \\
12 &  8 & 2.463 & +5.420 & +1.455 & 3.506 \\
12 & 10 & 2.487 & +5.636 & +1.455 & 3.490 \\
12 & 12 & 2.520 &        &        &       \\
11 &  6 & 2.514 & +5.525 & +1.515 & 3.491 \\
11 &  8 & 2.528 & +5.696 & +1.501 & 3.495 \\
11 & 10 & 2.542 & +5.871 & +1.496 & 3.539 \\
\multicolumn{2}{c}{Expt.} & 2.45(5) & (+)6.83(21) & 1.0355(3) & 9.1(4.5) \\
\multicolumn{6}{c}{INOY} \\
$\hbar\Omega$ & $N_{\rm max}$ & $r_{\rm p}$ & Q & $\mu$ & B(M1; $1^+\rightarrow 2^+$) \\
\hline
16 &  6 & 2.199 & +4.049 & +1.192 & 3.669 \\
16 &  8 & 2.241 & +4.306 & +1.207 & 3.669 \\
16 & 10 & 2.277 & +4.580 & +1.227 & 3.649 \\
16 & 12 & 2.317 & & & \\
15 &  6 & 2.241 & +4.242 & +1.238 & 3.681 \\
15 &  8 & 2.276 & +4.468 & +1.244 & 3.684 \\
15 & 10 & 2.305 & +4.710 & +1.257 & 3.667 \\
14 &  6 & 2.291 & +4.467 & +1.288 & 3.696 \\
14 &  8 & 2.318 & +4.660 & +1.286 & 3.702 \\
14 & 10 & 2.340 & +4.867 & +1.291 & 3.688 \\
\multicolumn{2}{c}{Expt.} & 2.45(5) & (+)6.83(21) & 1.0355(3) & 9.1(4.5) \\
\hline
\multicolumn{6}{c}{$^8$Li} \\
\multicolumn{6}{c}{CD-Bonn 2000} \\
$\hbar\Omega$ & $N_{\rm max}$ & $r_{\rm p}$ & Q & $\mu$ & B(M1; $1^+\rightarrow 2^+$) \\
\hline
12 &  6 & 2.139 & +2.588 & +1.238 & 4.454 \\
12 &  8 & 2.139 & +2.690 & +1.243 & 4.428 \\
12 & 10 & 2.145 & +2.784 & +1.241 & 4.393 \\
12 & 12 & 2.161 & & & \\
\multicolumn{2}{c}{Expt.} & 2.26(2) & +3.27(6) & +1.654 & 5.01(1.61) \\
\multicolumn{6}{c}{INOY} \\
$\hbar\Omega$ & $N_{\rm max}$ & $r_{\rm p}$ & Q & $\mu$ & B(M1; $1^+\rightarrow 2^+$) \\
\hline
16 &  6 & 1.938 & +2.279 & +1.469 & 4.635 \\
16 &  8 & 1.956 & +2.377 & +1.456 & 4.610 \\
16 & 10 & 1.972 & +2.477 & +1.439 & 4.578 \\
16 & 12 & 1.991 & & & \\
\multicolumn{2}{c}{Expt.} & 2.26(2) & +3.27(6) & +1.654 & 5.01(1.61) \\
    \end{tabular}
  \end{ruledtabular}
\end{table}

\section{\label{sec:overlap}Cluster form factors from {\it ab initio} wave functions}
\subsection{\label{subsec:NCSM_overlap}Overlap functions and spectroscopic factors obtained in the model space}

Detailed knowledge of nuclear structure is important for the
description of low-energy nuclear reactions.
As the first step in the application of the NCSM to low-energy nuclear reactions,
one needs to understand the cluster structure of the eigenstates.
This is achieved by calculating the channel cluster form factors
(or overlap integrals, overlap functions). The formalism for calculating
the channel cluster form factors from the NCSM wave functions was developed in Ref.~\cite{cluster}.
Here we just briefly repeat and adapt a part of the formalism relevant for our present
application.

We consider a composite system of $A$ nucleons, i.e. $^8$B, a nucleon projectile, here a proton,
and an $A-1$-nucleon target, i.e. $^7$Be.
Both nuclei are assumed to be described by eigenstates of the NCSM effective Hamiltonians
expanded in the HO basis with identical HO frequency and the same
(for the eigenstates of the same parity) or differing by one unit
of the HO excitation (for the eigenstates of opposite parity)
definitions of the model space. The target and the composite system is assumed to be described
by wave functions expanded in Slater determinant single-particle HO basis (that is obtained from
a calculation using a shell model code like Antoine).

Let us introduce a projectile-target wave function
\begin{eqnarray}\label{proj-targ_state_delta}
&&\langle\vec{\xi}_1 \ldots \vec{\xi}_{A-2} r^\prime \hat{r}
|\Phi_{(l\frac{1}{2})j;\alpha I_1}^{(A-1,1)J M};\delta_{r}\rangle
\nonumber \\
&=&\sum (j m I_1 M_1 | J M) (l m_l \textstyle{\frac{1}{2}} m_s | j m)
\frac{\delta(r-r^\prime)}{r r^\prime}
\nonumber \\
&\times&
Y_{l m_l}(\hat{r}) \chi_{m_s}
\langle \vec{\xi}_1 \ldots \vec{\xi}_{A-2} | A-1 \alpha I_1 M_1\rangle \; ,
\end{eqnarray}
where
$\langle \vec{\xi}_1 \ldots \vec{\xi}_{A-2} | A-1 \alpha I_1 M_1\rangle$
and
$\chi_{m_s}$ are the target and the nucleon wave function, respectively.
Here, $l$ is the channel relative orbital angular momentum, $\vec{\xi}$ are the target
Jacobi coordinates and
$\vec{r}=\left[\frac{1}{A-1}
      \left(\vec{r}_1+\vec{r}_2 + \ldots+ \vec{r}_{A-1}\right)-\vec{r}_{A}\right]$
describes the relative distance between the nucleon and the center of mass of the target.
The spin and isospin coordinates were omitted for simplicity.

The channel cluster form factor is then defined by
\begin{equation}\label{cluster_form_factor}
g^{A\lambda J}_{(l\frac{1}{2})j;A-1 \alpha I_1}(r)=
\langle A \lambda J |{\cal A}\Phi_{(l\frac{1}{2})j;\alpha I_1}^{(A-1,1)J};
\delta_{r}\rangle \; ,
\end{equation}
with ${\cal A}$ the antisymmetrizer and $|A\lambda J\rangle$ an eigenstate 
of the $A$-nucleon composite system (here $^8$B). It can be calculated from the
NCSM eigenstates obtained in the Slater-determinant basis from a
reduced matrix element of the creation operator:
\begin{eqnarray}\label{single-nucleon}
\langle A \lambda J|&{\cal A}& \Phi_{(l \textstyle{\frac{1}{2}},j);\alpha I_1}^{(A-1,1) J};
\delta_{r}\rangle
= \sum_n R_{nl}(r)
\nonumber \\
&\times&
\frac{1}{\langle nl00l|00nll\rangle_{\frac{1}{A-1}}} \frac{1}{\hat{J}}
(-1)^{I_1-J-j}
\nonumber \\
&\times&
\; _{\rm SD}\langle A\lambda J||a^\dagger_{nlj}||A-1\alpha I_1\rangle_{\rm SD} \;
\; .
\end{eqnarray}
The subscript SD refers to the fact that these states were obtained
in the Slater determinant basis, i.e. by using
a shell model code, and, consequently, contain spurious center-of-mass components.
A general HO bracket, which value is simply given by
\begin{equation}\label{cm_ho_br}
\langle nl00l|00nll\rangle_{\frac{1}{A-1}} = (-1)^l
\left(\frac{A-1}{A}\right)^{\frac{2n+l}{2}}
\; ,
\end{equation}
then appears in Eq. (\ref{single-nucleon}) in order to remove these components.
The $R_{nl}(r)$ in Eq. (\ref{single-nucleon}) is the radial HO
wave function with the oscillator length parameter $b=\sqrt{\frac{\hbar}{\frac{A-1}{A}m\Omega}}$,
where $m$ is the nucleon mass.

A conventional spectroscopic factor is obtained by integrating the square of the cluster form
factor:
\begin{equation}\label{spec_fac}
S^{A\lambda J}_{(l\frac{1}{2})j;A-1 \alpha I_1}=
\int dr r^2
|g^{A\lambda J}_{(l\frac{1}{2})j;A-1 \alpha I_1}(r)|^2
\; .
\end{equation}
We calculated the $\langle ^8$B$|^7$Be+p$\rangle$ channel cluster
form factors and the spectroscopic factors for the bound states of
$^8$B and $^7$Be from the NCSM wave functions obtained in model
spaces up to $N_{\rm max}=10$ ($10\hbar\Omega$) in a wide HO
frequency range and using both the CD-Bonn and the INOY NN
potentials. The most important channels are the $p$-waves, $l=1$,
with the proton in the $j=3/2$ and $j=1/2$ states. Our selected
spectroscopic factor results are summarized in
Table~\ref{tab:B8specfac}. It should be noted a very weak dependence
of the spectroscopic factors on either the basis size or the HO
frequency or the NN potential.

\begin{table}[hbtp]
  \caption{The $\langle ^8$B$ (2^+_{\rm gs}) | ^7$Be$ (I_1^\pi) + $p$(l,j)\rangle$ 
spectroscopic factors obtained within the NCSM
for different HO frequencies and model spaces for the CD-Bonn 2000 and INOY NN potentials.
  \label{tab:B8specfac}}
  \begin{ruledtabular}
    \begin{tabular}{ccccc}
\multicolumn{2}{c}{CD-Bonn 2000} & \multicolumn{3}{c}{$^7$Be+p $I_1^\pi (l,j)$} \\
$\hbar\Omega$ [MeV] & $N_{\rm max}$  &  $\frac{3}{2}^- (1,\frac{3}{2})$ &  $\frac{3}{2}^- (1,\frac{1}{2})$
&  $\frac{1}{2}^- (1,\frac{3}{2})$ \\
\hline
11 &  6 & 0.977 & 0.120 & 0.285 \\
11 &  8 & 0.967 & 0.116 & 0.280 \\
11 & 10 & 0.959 & 0.111 & 0.275 \\
12 &  6 & 0.978 & 0.107 & 0.287 \\
12 &  8 & 0.969 & 0.103 & 0.281 \\
12 & 10 & 0.960 & 0.102 & 0.276 \\
14 &  6 & 0.979 & 0.086 & 0.288 \\
14 &  8 & 0.967 & 0.085 & 0.284 \\
14 & 10 & 0.958 & 0.085 & 0.280 \\
\hline
\multicolumn{2}{c}{INOY} & \multicolumn{3}{c}{$^7$Be+p $I_1^\pi (l,j)$} \\
$\hbar\Omega$ [MeV] & $N_{\rm max}$  &  $\frac{3}{2}^- (1,\frac{3}{2})$ &  $\frac{3}{2}^- (1,\frac{1}{2})$
&  $\frac{1}{2}^- (1,\frac{3}{2})$ \\
\hline
14 &  6 & 0.987 & 0.074 & 0.283 \\
14 &  8 & 0.976 & 0.072 & 0.277 \\
14 & 10 & 0.966 & 0.072 & 0.272 \\
16 &  6 & 0.988 & 0.060 & 0.281 \\
16 &  8 & 0.977 & 0.061 & 0.276 \\
16 & 10 & 0.965 & 0.063 & 0.271 \\
    \end{tabular}
  \end{ruledtabular}
\end{table}

We note that the $\langle ^8$Li$|^7$Li$+$n$\rangle$ spectroscopic factors were
obtained from $^7$Li(d,p)$^8$Li stripping measurements through a distorted-wave 
Born approximation (DWBA) analysis
\cite{Macfarlane,Schiffer}. The NCSM $\langle ^8$Li$|^7$Li$+$n$\rangle$ and
$\langle ^8$B$|^7$Be$+$p$\rangle$ spectroscopic factors are almost identical
and overestimate those extracted in Ref.~\cite{Macfarlane,Schiffer} with the ground-state
values of about 0.9. At the same time, the NCSM spectroscopic factors are in good agreement
with those obtained in R-matrix analysis by Barker \cite{Barker95}. Also, they
are consistent with the spectroscopic factors from the microscopic three-cluster model
\cite{D04} as well as with the Cohen-Kurath spectroscopic factors \cite{CK67}
with no center-of-mass correction.  

\subsection{\label{subsec:corr_overlap}Correction of the overlap-function asymptotics}

The dominant $p$-wave $j=3/2$ overlap integral for the $^8$B and $^7$Be ground states
obtained using the CD-Bonn NN potential in the $10\hbar\Omega$ model space
and the HO frequency of $\hbar\Omega=12$ MeV
is presented in 
Fig. \ref{B8_Be7+p_overlap_12_3} by the full line. Despite the fact
that a very large basis was employed in the present calculation, it
is apparent that the overlap integral is nearly zero at about 10 fm.
This is a consequence of the HO basis asymptotics. 
A separate issue is the incorrect threshold placements obtained
from the high precision NN potentials that we employ.
The proton
capture on $^7$Be to the weakly bound ground state of $^8$B
associated dominantly by the $E1$ radiation is a peripheral process.
Consequently, the overlap integral with an incorrect asymptotic
behavior cannot be used to calculate the S-factor.

\begin{figure}[hbtp]
  \includegraphics*[width=0.9\columnwidth]
   {overlap_plot_fit.dat_B8_Be7cdb2k_10.12_l1j3Ia3_whitt.eps}
  \caption{Overlap integral, $rg(r)$, for the ground state of $^8$B with the ground
state of $^7$Be plus proton as a function of separation between the $^7$Be
and the proton. The $p$-wave channel with $j=3/2$ is shown.
The full line represents the NCSM result obtained using the CD-Bonn NN potential
in the $10\hbar\Omega$ model space and the HO frequency of $\hbar\Omega=12$ MeV.
The dashed line represents a corrected overlap obtained from a Woods-Saxon
potential whose parameters were fit to the NCSM overlap up to 4.0 fm under the
constraint to reproduce the experimental separation energy.
The dotted line represents a corrected overlap obtained by a direct matching 
of the NCSM overlap and the Whittaker function.%
  \label{B8_Be7+p_overlap_12_3}}
\end{figure}
\begin{figure}[hbtp]
  \includegraphics*[width=0.9\columnwidth]
{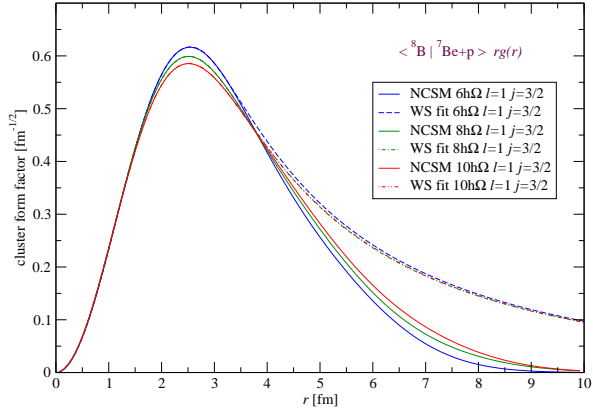}
  \caption{Same as in Fig.~\ref{B8_Be7+p_overlap_12_3}, but for a comparison
of results obtained using model spaces of $6,8,10\hbar\Omega$. Only corrections 
by the WS potential fit are shown.
The fitting ranges up to 3.6, 3.8 and 4.0 fm were used, respectively.
  \label{B8_Be7+p_overlap_cdb2k_6810_12}}
\end{figure}
\begin{figure}[hbtp]
  \includegraphics*[width=0.9\columnwidth]
   {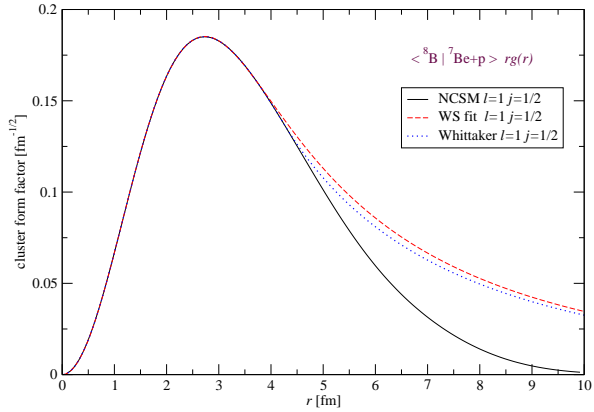}
  \caption{Same as in Fig.~\ref{B8_Be7+p_overlap_12_3}, but for
the $p$-wave channel with $j=1/2$.%
  \label{B8_Be7+p_overlap_12_1}}
\end{figure}

It is our expectation, however, that the interior part of the overlap integral
as obtained from our large-basis NCSM calculation is realistic. 
This is supported by the fact that the overlap integrals obtained using the CD-Bonn 2000
and the INOY NN potentials are similar despite the fact that the two potentials
predict different binding energies and thresholds. Further, we calculated
overlap integrals using a Hamiltonian that included the three-nucleon interaction.
In those calculations, for technical reasons limited to small model spaces up 
to $6\hbar\Omega$, we obtained binding energies closer to experiment.
Still, the interior part of the overlap integrals was almost the same as that 
calculated here from the CD-Bonn 2000 in the same model space and identical HO frequency. 
It is then straightforward
to correct the asymptotic behavior of the overlap integral and fix in this way
both the HO basis and the wrong threshold placement issues at the same time. 
One possibility we explored utilizes solutions of a Woods-Saxon potential. 
In particular, we performed
a least-square fit of a WS potential solution to the interior of the
NCSM overlap in the range of $0-4$ fm. The WS potential parameters
were varied in the fit under the constraint that the experimental
separation energy of $^7$Be+p, i.e. 0.137 MeV, was reproduced. In this way we obtain a perfect
fit to the interior of the overlap integral and a correct asymptotic behavior
at the same time. The resulting overlap function is presented in Fig. \ref{B8_Be7+p_overlap_12_3}
by the dashed line. 

Another possibility is a direct matching of logarithmic derivatives of the NCSM overlap integral
and the Whittaker function:
\begin{equation}\label{logdermatch}
\frac{d}{dr}ln(rg_{lj}(r))=\frac{d}{dr}ln(C_{lj} W_{-\eta,l+1/2}(2k_0r)) \; ,
\end{equation}
where $\eta$ is the Sommerfeld parameter, $k_0=\sqrt{2\mu E_0}/\hbar$ with $\mu$ the reduced mass
and $E_0$ the separation energy (here $E_0=0.137$~MeV). 
The NCSM overlap integral $g_{lj}(r)$ is defined in Eq.~(\ref{cluster_form_factor}), and, according to
Eq.~(\ref{single-nucleon}), it is expanded in terms of the HO radial wave functions, i.e. 
$rg_{lj}(r)=\sum_n s_{nlj}{\cal R}_{nl}(r)$ with ${\cal R}_{nl}(r)=rR_{nl}(r)$. 
The spectroscopic factor (\ref{spec_fac}) is then given by $S_{lj}=\sum_n s_{nlj}^2$.
For simplicity, we suppressed
all quantum numbers except the orbital momentum $l$, the total proton angular momentum $j$ 
and the radial HO quantum number $n$. With the help of 
\begin{eqnarray}\label{R_deriv}
\frac{d}{dr}{\cal R}_{nl}(r)&=& \left(\frac{2n+l+1}{r}-\frac{r}{b^2}\right){\cal R}_{nl}(r)
\nonumber \\
&&
-\sqrt{n(n+l+\textstyle{\frac{1}{2}})}\frac{2}{r}{\cal R}_{n-1 l}(r) \; ,
\end{eqnarray}
it is straightforward to calculate the derivative of $rg_{lj}(r)$.
Since asymptotic normalization constant (ANC) $C_{lj}$
cancels out, there is a unique solution of Eq.~(\ref{logdermatch}) at $r=R_m$.
For the discussed overlap presented in Fig.~\ref{B8_Be7+p_overlap_12_3}, we found $R_m=4.05$~fm.
Finally, by matching the value of the NCSM overlap and the Whittaker function at $R_m$, 
we determine the ANC. 
The corrected overlap using the Whittaker function matching is shown in Fig.~\ref{B8_Be7+p_overlap_12_3}
by a dotted line. In general, we observe that the approach using the WS fit leads to deviations from the
original NCSM overlap starting at a smaller radius. In addition, the WS solution fit introduces
an intermediate range from about 4 fm to about 6 fm, where the corrected overlap deviates
from both the original NCSM overlap and the Whittaker function. Perhaps, this is a more realistic
situation compared to the direct Whittaker function matching. In any case, by considering the two alternative
procedures we are in a better position to estimate uncertainties in our S-factor results. This
is discussed later.

In the end, we re-scale the corrected overlap functions to preserve the original
NCSM spectroscopic factor (given in Table \ref{tab:B8specfac}).
In general, we observe a faster convergence of the spectroscopic
factors than that of the overlap functions. With increasing basis size ($N_{\rm max}$), 
the maximum of the overlap functions decreases while the tail extends
with the integral of the square approximately conserved. This is demonstrated 
in Fig.~\ref{B8_Be7+p_overlap_cdb2k_6810_12}, see further discussion below.
The corrected
overlap function should represent the infinite space result. By re-scaling
a corrected overlap function obtained at a finite $N_{\rm max}$, we approach
faster the infinite space result. 

The same procedure is applied to other relevant channels.
In Fig.~\ref{B8_Be7+p_overlap_12_1}, we present the NCSM overlap integral and its corrected
form for the other $p$-wave channel with proton in $j=\frac{1}{2}$ and $^8$B and $^7$Be
in their ground states.
Obviously, the WS parameters as well as the matching radii $R_m$ 
obtained for the two channels are different (in the case of the present
$j=\frac{1}{2}$ channel we found $R_m=4.23$~fm).
In Table~\ref{tab:WSparam}, we show the fitted WS potential parameters
obtained in the two discussed cases together with parameters corresponding to a $p$-wave
channel with $^7$Be in the first excited state. In this last case, the separation energy
is $E_0=0.57$ MeV.
We use the definition of the WS potential as given, e.g. in Eqs. (5-7) of Ref.~\cite{Radcap}.
Typically, the central potential parameters
$R_0$, $a_0$ are well constrained in the fit, while the spin-orbit potential parameters
are obtained with some uncertainty. Their values then exhibit more variation from channel
to channel. The strength of the central potential $V_0$ was re-adjusted at every step
during the fit to reproduce the experimental separation energy. We note that parameter values
in Table~\ref{tab:WSparam} are rounded compared to the results from our fitting procedure. 
One needs to fine-tune, e.g. the $V_0$, to reproduce accurately the respective separation 
energy in order to use the parameters independently.

It should be noted that the Woods-Saxon
potential here is just a tool to represent the interior part of the NCSM overlap integral
as accurately as possible and correct its asymptotic form at the same time. The range used
in the least-square fit is not arbitrary and varies from channel to channel. The aim is to
use as large range as possible, while at the same time preserve the NCSM overlap integral
as accurately as possible in that range. A combination of eye-guided evaluation
with a quantitative condition of minimizing the least-square function per fitted point
helps to determine the largest possible range. Finally, let us repeat that the alternative
procedure of the direct Whittaker function matching is completely unique. 

\begin{table}[hbtp]
  \caption{Parameters of the Woods-Saxon potentials obtained in the fits to the
interior part of the NCSM $\langle ^8$B$ (2^+_{\rm gs}) | ^7$Be$ (I_1^\pi) + $p$(l,j)\rangle (r)$
overlap functions under the constraint to reproduce experimental
separation energies. The $p$-wave channels for the $^7$Be ground and the first excited state
are shown. Results for the $10\hbar\Omega$ model space and HO frequencies of
$\hbar\Omega=12(14)$ MeV for the CD-Bonn 2000 (INOY) NN potential. Woods-Saxon potential
parameters from Ref.\cite{Esbensen} (with slight modification of $V_0$)
that we use for the scattering states are also shown.
  \label{tab:WSparam}}
  \begin{ruledtabular}
    \begin{tabular}{cccccccc}
\multicolumn{8}{c}{CD-Bonn 2000 $10\hbar\Omega$ $\hbar\Omega=12$ MeV} \\
$I_1^\pi (l,j)$ & $V_0$ & $R_0$ & $a_0$ & $V_{\rm ls}$ & $R_{\rm ls}$ & $a_{\rm ls}$ & $R_{\rm C}$ \\
\hline
$\frac{3}{2}^- (1,\frac{3}{2})$ & -51.037 & 2.198 & 0.602 & -9.719 & 2.964 & 0.279 & 2.198 \\
$\frac{3}{2}^- (1,\frac{1}{2})$ & -45.406 & 2.613 & 0.631 & -8.414 & 2.243 & 0.366 & 2.613 \\
$\frac{1}{2}^- (1,\frac{3}{2})$ & -49.814 & 2.235 & 0.553 & -17.024& 3.080 & 0.338 & 2.235 \\
\hline
\multicolumn{8}{c}{INOY $10\hbar\Omega$ $\hbar\Omega=14$ MeV} \\
$I_1^\pi (l,j)$ & $V_0$ & $R_0$ & $a_0$ & $V_{\rm ls}$ & $R_{\rm ls}$ & $a_{\rm ls}$ & $R_{\rm C}$ \\
\hline
$\frac{3}{2}^- (1,\frac{3}{2})$ & -58.836 & 2.052 & 0.561 & -7.518 & 2.768 & 0.253 & 2.052 \\
$\frac{3}{2}^- (1,\frac{1}{2})$ & -55.924 & 2.470 & 0.580 & -17.454& 2.027 & 0.429 & 2.470 \\
$\frac{1}{2}^- (1,\frac{3}{2})$ & -44.300 & 2.455 & 0.509 & -13.325& 1.011 & 0.347 & 2.455 \\
\hline
\multicolumn{8}{c}{scattering state} \\
  & $V_0$ & $R_0$ & $a_0$ & $V_{\rm ls}$ & $R_{\rm ls}$ & $a_{\rm ls}$ & $R_{\rm C}$ \\
\hline
  & -42.2 & 2.391 & 0.52 & -9.244 & 2.391 & 0.52 & 2.391 \\
    \end{tabular}
  \end{ruledtabular}
\end{table}

The corrected overlap integrals then serve as the
input for the momentum distribution and the S-factor calculations as described in the
following sections.

\begin{figure}[hbtp]
  \includegraphics*[width=0.9\columnwidth]
{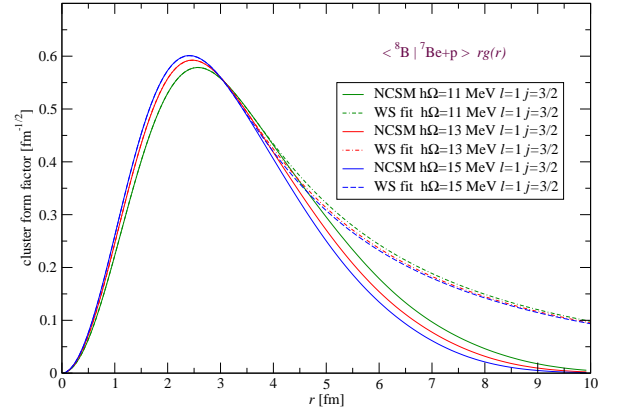}
  \caption{Same as in Fig.~\ref{B8_Be7+p_overlap_12_3}, but for a comparison
of results obtained using the HO frequencies of $\hbar\Omega=11,13,15$ MeV.
Only corrections by the WS potential fit are shown.
The fitting ranges up to 4.0, 3.8 and 3.7 fm were used, respectively.
  \label{B8_Be7+p_overlap_cdb2k_10_111315}}
\end{figure}

\begin{figure}[hbtp]
  \includegraphics*[width=0.9\columnwidth]
{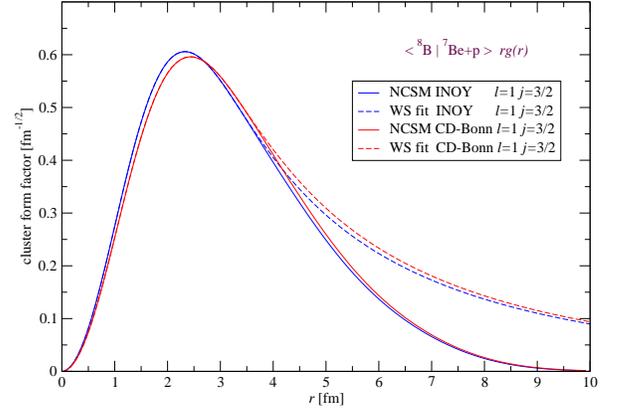}
  \caption{Same as in Fig.~\ref{B8_Be7+p_overlap_12_3}, but using
the HO frequency of $\hbar\Omega=14$ MeV with results obtained by
the CD-Bonn 2000 and INOY NN potentials shown for a comparison.
Only corrections by the WS potential fit are shown.
The fitting ranges up to 3.75 and 3.7 fm were used, respectively.
  \label{B8_Be7+p_overlap_cdb2k_INOY_14}}
\end{figure}

The basis size dependence of the overlap integrals both the original
NCSM and the corrected ones (using the WS solution fit procedure)
can be judged from
Fig.~\ref{B8_Be7+p_overlap_cdb2k_6810_12}, where we compare results
obtained in model spaces from $6\hbar\Omega$ to $10\hbar\Omega$
using the CD-Bonn 2000 NN potential and a fixed HO frequency of
$\hbar\Omega=12$ MeV. With an increase of the basis, the maximum of
the overlap decreases and its tail extends to larger distances. The
corresponding spectroscopic factors differ very little, see
Table~\ref{tab:B8specfac}. The change from $8\hbar\Omega$ to
$10\hbar\Omega$ is smaller than that between $6\hbar\Omega$ and
$8\hbar\Omega$, a sign of convergence. In
Fig.~\ref{B8_Be7+p_overlap_cdb2k_10_111315}, we then compare the
overlap integrals, both original and corrected, obtained in the
$10\hbar\Omega$ model space using three different HO frequencies.
With a decrease of the HO frequency, the maximum of the overlap
decreases and shifts to a larger distance. At the same time, the
tail extends to larger distances as well. Finally, in
Fig.~\ref{B8_Be7+p_overlap_cdb2k_INOY_14} we compare overlap
integrals obtained using the CD-Bonn 2000 and the INOY NN potentials
in the same $10\hbar\Omega$ model spaces and the same HO frequency
of $\hbar\Omega=14$ MeV. While the spectroscopic factors are almost
identical, the CD-Bonn overlap integral has a smaller maximum
shifted to a larger distance and also extends further. The
parameters of the WS potential that produce the INOY
corrected overlap integral are given in Table~\ref{tab:WSparam}. The
impact of these discussed dependencies on the
$^7$Be(p,$\gamma$)$^8$B S-factor is discussed in
Sect.~\ref{sec:sfactor}.


\section{\label{sec:momdis}Momentum distributions}

The stripping reaction $(^7{\rm Be}+p)+A\longrightarrow \ ^7{\rm
Be}+X$ cross section, for a specified final state of the core ($^7{\rm Be}$), is
given by \cite{HM85} (for more details, see Ref. \cite{BH04})
\begin{eqnarray}
&&\frac{d\sigma_{\mathrm{str}}}{d^{3}k_{c}}=\frac{1}{\left(  2\pi\right)  ^{3}%
}\frac{1}{2l_b+1}\sum_{m_b}\int d^{2}b_{p}\left[  1-\left\vert
S_{p}\left( b_{p}\right)  \right\vert ^{2}\right]
\nonumber\\
&\times& \left\vert \int d^{3}r\ e^{-i\mathbf{k}%
_{c}\mathbf{.r}}S_{c}\left(  b_{c}\right)  
g^{A\lambda J_b}_{(l_b\frac{1}{2})j_b;A-1 \alpha I_1}(r) Y_{l_bm_b}(\hat{\bf r})
\right\vert ^{2}, \label{sknock}%
\end{eqnarray}
where $\mathbf{r}\equiv\left( r, \theta, \phi \right)\equiv \left(
\rho,z,\phi\right) =\mathbf{r}_{p}-\mathbf{r}_{c}$. 
$S_{c}$ ($S_{p}$) are the S-matrices for the
core($^7{\rm Be}$)+target and the proton+target scattering,
respectively. The overlap integral
$g^{A\lambda J_b}_{(l_b\frac{1}{2})j_b;A-1 \alpha I_1}(r)$ 
describing the bound state of the $(^7{\rm Be+p})$ subsystem
is defined as in Eq. (\ref{cluster_form_factor}) with its asymptotic 
tail corrected as described in Sect.~\ref{subsec:corr_overlap}. 

The cross sections for the longitudinal momentum distributions are
obtained by
integrating Eq. (\ref{sknock}) over the transverse component of $\mathbf{k}%
_{c}$, i.e.
\begin{eqnarray}
& &\frac{d\sigma_{\mathrm{str}}}{dk_z}    = \frac{1}{2\pi}\frac{1}%
{2l_b+1}\sum_{m_b}\int d^{2}b_{p}\left[  1-\left\vert
S_{p}\left(
b_{p}\right)  \right\vert ^{2}\right]\nonumber\\
&\times&  
\int d^{2}%
\rho\left\vert S_{c} \left(  b_{c}\right)  \right\vert ^{2}
\nonumber\\
&  \times& \left\vert
\int_{-\infty}^{\infty}dz\exp\left[ -ik_{z}z\right]
g^{A\lambda J_b}_{(l_b\frac{1}{2})j_b;A-1 \alpha I_1}(r) Y_{l_bm_b}(\hat{\bf r})
\right\vert ^{2}, \nonumber \\
\label{strL}%
\end{eqnarray}
where $k_z$ represents the longitudinal component of ${\bf k}_c$.

For the transverse momentum distribution in cylindrical coordinates
$k_\bot\equiv \mathbf{k}_{c}^{\perp}=\sqrt{k_{x}^{2}+k_{y}^{2}}$
\begin{eqnarray}
&&\frac{d\sigma_{\mathrm{str}}}{d^{2}k_\bot}   =
\frac{1}{(2\pi)^2}\frac {1}{2l_b+1}\ \int_{0}^{\infty}d^{2}b_{p}\
\left[ 1-\left\vert S_{p}\left( b_{p}\right)  \right\vert
^{2}\right]
 \nonumber\\
&  \times&
\sum_{m_b}\  \int_{-\infty}^{\infty}dz\ 
\left\vert \int d^{2}%
\rho\ \exp\left(  -i\mathbf{k}_{c}^{\perp}\mathbf{.\mbox{\boldmath$\rho$}}%
\right)  S_{c}\left(  b_{c}\right)  \right.
 \nonumber\\
&  \times&
\left.
g^{A\lambda J_b}_{(l_b\frac{1}{2})j_b;A-1 \alpha I_1}(r) Y_{l_bm_b}(\hat{\bf r})
\right\vert ^{2}. \nonumber\\
\label{strT}%
\end{eqnarray}

It is also convenient to describe the transverse momentum
distributions in terms of the projection onto one of the Cartesian
components of the transverse momentum, i.e.%
\begin{equation}
\frac{d\sigma_{\mathrm{str}}}{dk_{y}}=\int dk_{x}\ \frac{d\sigma
_{\mathrm{str}}}{d^{2}k_\bot}\left(  k_{x},k_{y}\right)  \ . \label{sigtx}%
\end{equation}

The total stripping cross section can be obtained by integrating
either Eq. (\ref{strL}) or Eq. (\ref{strT}). The integrals in Eqs.
(\ref{strL}), (\ref{strT}), and for the total cross sections, are
done numerically with use of Gaussian expansions of the S-matrices,
as explained in Ref. \cite{BH04}.

The S-matrices have been obtained using the optical limit of the
Glauber theory, i.e.
\begin{equation}
S_i\left(  b\right)  =\exp\left[  \frac{i}
{k_{NN}}\int_{0}^{\infty}dq\ q \rho_{i}\left(  q\right)
\rho_{t}\left( q\right)  f_{NN}\left( q\right)  J_{0}\left(
qb\right)
\right]  , \label{intro3}%
\end{equation} where $\rho_{i,t}\left(  q\right)  $ is the
Fourier transform of the nuclear densities of the incident particle
($i={\rm p},\ ^7{\rm Be}$) and target (t), and $f_{NN}\left(
q\right) $ is the high-energy nucleon-nucleon scattering amplitude
at forward angles, which can be parametrized
by \cite{ray79}%
\begin{equation}
f_{NN}\left(  q\right)  =\frac{k_{NN}}{4\pi}\sigma_{NN}\left(  i
+\alpha _{NN}\right)  \exp\left(  -\beta_{NN}q^{2}\right)  \ .
\end{equation}

In this equation $\sigma_{NN}$, $\alpha_{NN},$ and $\beta_{NN}$ are
parameters which fit the high-energy nucleon-nucleon scattering at
forward angles \cite{ray79}. We use the values from Ref.
\cite{RHB91} which have been extended to collisions at lower
energies and corrected for isospin-average. The quantities
$\rho_{i}\left( q\right) $ ($i={\rm p},\ ^7{\rm Be}$) and $\
\rho_{t}\left( q\right) $ are calculated from radial density
distributions taken to be of Gaussian shapes, adjusted to reproduce
the rms radius of the proton, $^7{\rm Be}$, $^9{\rm Be}$ and
$^{12}{\rm C}$, respectively.

The Coulomb part of the eikonal phase is included according the
prescription described in details in Ref. \cite{BH04}.

\begin{figure}[hbtp]
  \includegraphics*[width=0.9\columnwidth]
   {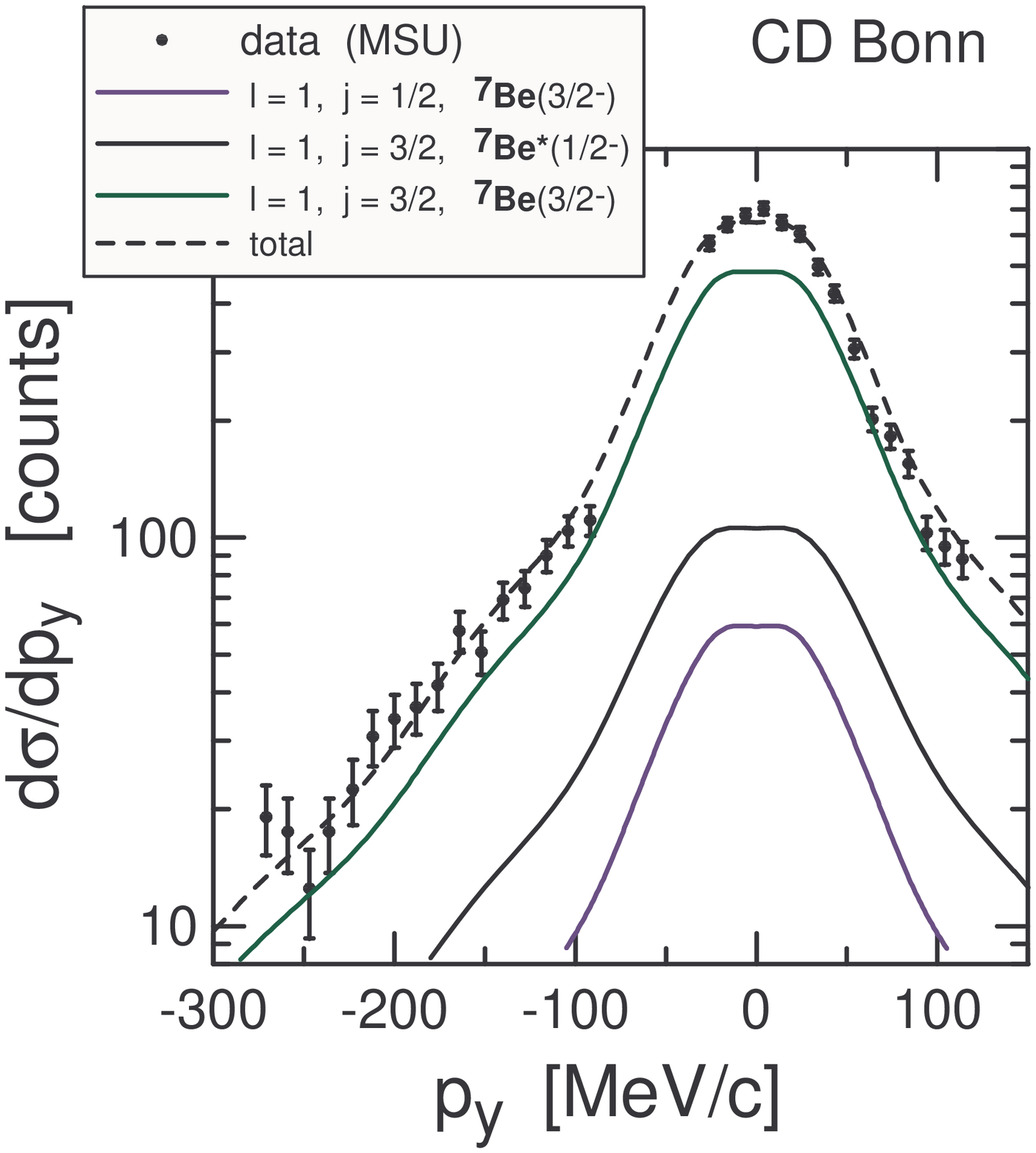}
\caption{Inclusive transverse-momentum distribution for the residue
in the $^{9}$Be($^{8}$B,$^{7}$Be)X reaction measured at 41
MeV/nucleon \cite{kel96}. The theoretical calculations are based on
Eqs. \ref{strT} and \ref{sigtx} and using the CD-Bonn interaction
for the $^8{\rm B}$ NCSM wave function. The dotted curve is the sum
of the individual contributions (full drawn).  In order of
increasing magnitude, the full curves correspond to $l = 1, j = 1/2;
\ ^7{\rm Be}(3/2^-)$, $l = 1, j = 3/2;  \ ^7{\rm Be}^*(1/2^-)$, and
$l = 1, j = 3/2;  \ ^7{\rm Be}(3/2^-)$ states, respectively. The
angular resolution in the experiment broadens the data by
approximately 4\%. This has not been included in the theoretical
curves.
  \label{MSU_cdb2k}}
\end{figure}

\begin{figure}[hbtp]
  \includegraphics*[width=0.9\columnwidth]
   {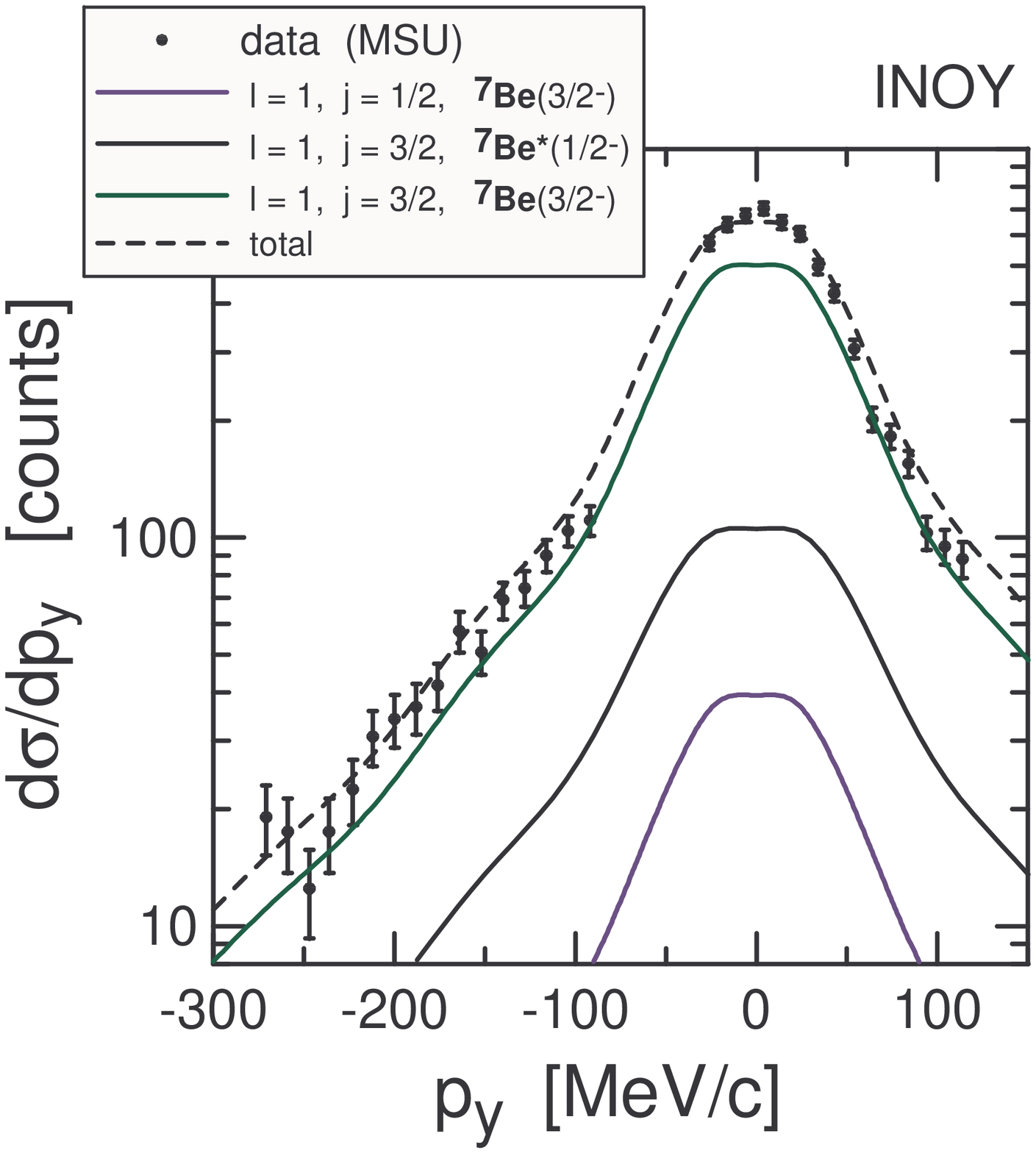}
\caption{Same as in Fig. \ref{MSU_cdb2k}, but using the INOY
interaction for the $^8{\rm B}$ NCSM wave function.
  \label{MSU_INOY}}
\end{figure}

As seen from Eqs.~(\ref{strL}) and (\ref{strT}), the longitudinal and
transverse momentum distributions, as described above, is a direct
test of the $^8{\rm B}$ ground state wave function calculated with
the NCSM. We do this by comparing to two sets of experimental data.
First we compare to an experiment performed at MSU for the
transverse momentum distributions of $^7{\rm Be}$ fragments from the
reaction $^8{\rm B}+^9{\rm Be}$ at 41 MeV/nucleon \cite{kel96}. In
spite of the absence of gamma-ray coincidence data, this is a
favorable case because the ground-state cross section dominates and
because the approximately 15\% branch to the excited level also has
$l$=1 and has an almost identical shape. Because the data is given
in arbitrary units, we have multiplied  the contributions for the
momentum distributions from each angular momentum channel 
by the same factor so that
their sum reproduces the maximum of the experimental distribution.
Thus, by comparing our results to this experiment we are testing the
relative ratios between the $^8{\rm B}$  spectroscopic factors
obtained with the NCSM. Figures \ref{MSU_cdb2k} and \ref{MSU_INOY}
shows our calculations with the CD-Bonn and INOY interactions,
respectively.
The WS solution fit procedure was employed
to correct the asymptotics of the NCSM overlap functions.
We used the $10\hbar\Omega$ model space and the optimal
HO frequencies of $\hbar\Omega=12$ MeV and $\hbar\Omega=16$ MeV
as determined from the ground-state-energy dependencies for the CD-Bonn
and INOY potentials, respectively. 
In both cases one notices that our calculations are in
excellent agreement with the data, which they reproduce over two
orders of magnitude in cross section.

From Figs. \ref{MSU_cdb2k} and \ref{MSU_INOY} it is difficult to
judge the quality of the CD-Bonn and the INOY interaction in
reproducing the momentum distributions. Except from a noticeable
visual difference seen for the magnitude of individual contributions
for the 3 distinct $^8{\rm B}$  states, the shape of the total
distributions is essentially the same for the two interactions.


\begin{figure}[hbtp]
  \includegraphics*[width=0.9\columnwidth]
   {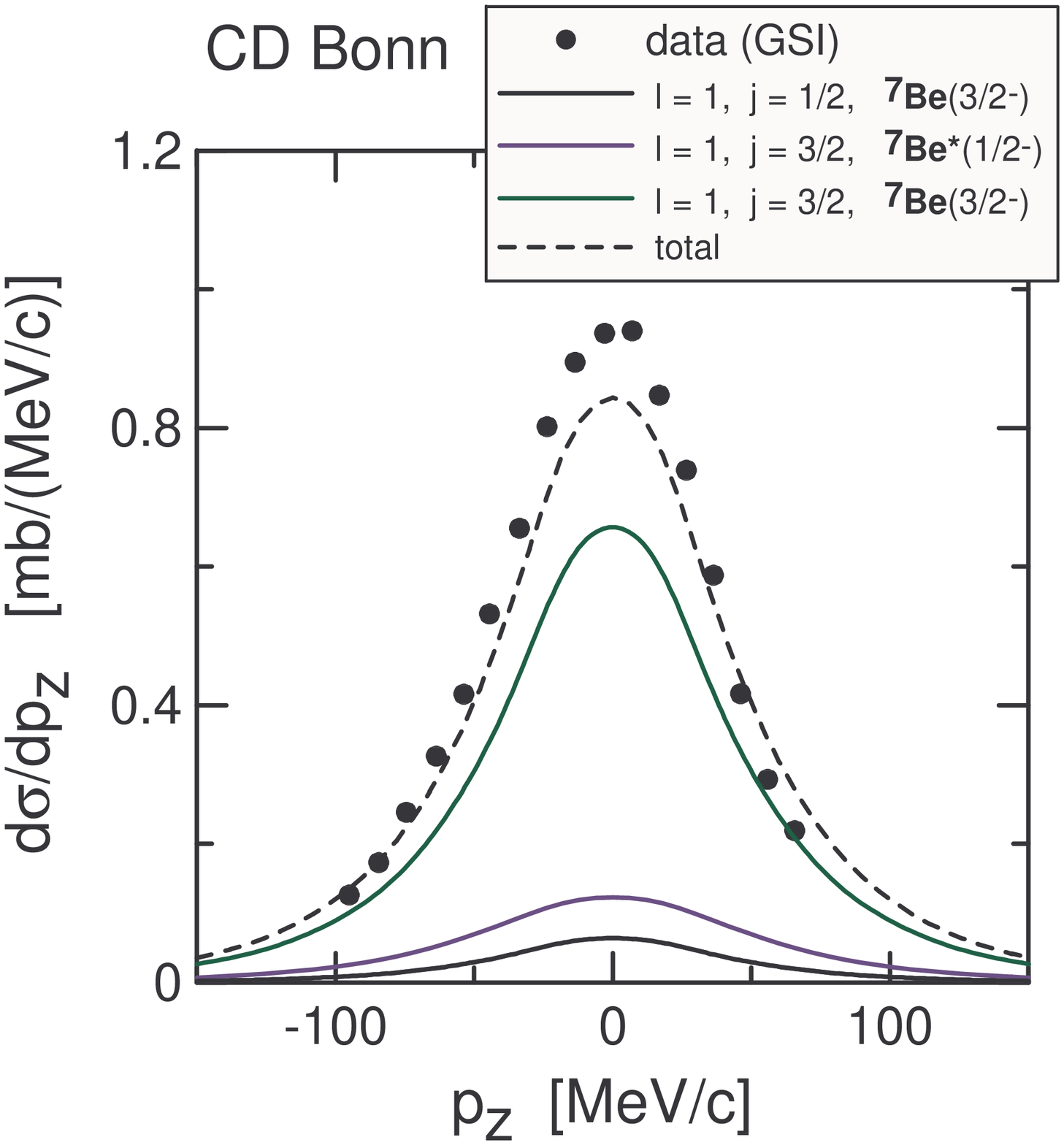}
  \caption{Inclusive parallel momentum distribution of $^7$Be
fragments from the reaction $^8$B+$^{12}$C$\longrightarrow^7$Be+$X$
at 936 MeV/nucleon. The full drawn curves represent theoretical
calculations using the NCSM with the CD-Bonn 2000 interaction. The
sum of the three theoretical contributions is shown by a
dashed-line.
  \label{GSI_cdb2k}}
\end{figure}

We perform another comparison, this time for the longitudinal
momentum distributions, using Eq.~(\ref{strL}). The data are taken
from a GSI experiment \cite{Gil02} which measured gamma rays in
coincidence with $^7{\rm Be}$ residues and separated the cross
sections to the three final states in the reaction $^8{\rm B}+^{12}{\rm
C}$ at 936 MeV/nucleon. Figure \ref{GSI_cdb2k} shows the data,
which are in reasonably good agreement with calculations. The full
drawn curves represent theoretical calculations using the NCSM with
the CD-Bonn 2000 interaction. The sum of the three theoretical
contributions is shown by a dashed-line. The experimental cross
section is $\sigma_{-p}=94\pm 9$ mb, whereas the theoretical one,
with NCSM wave functions and the CD-Bonn 2000 interaction, is 99.66
mb, in excellent agreement with the experimental one. The
theoretical total momentum distribution for this reaction is wider
than the experimental data. The experimental FWHM (full width at
half maximum) is $95\pm 5$ MeV/c if fitted by a single Gaussian. The
theoretical one has a longer tail than the experimental one. It can
be fitted by two Gaussians. The width of the narrower Gaussian is 96
MeV/c, also in good agreement with the experimental data. The reason
for the longer tail in the theoretical momentum distribution is not
well known, although a possible explanation is that the comparison
between experiment and theory, as described in Ref.~\cite{Gil02},
involves a folding with the experimental resolution and a scaling to
match the amplitude of the experimental spectrum, which has not been
done here.

\begin{figure}[hbtp]
  \includegraphics*[width=0.9\columnwidth]
   {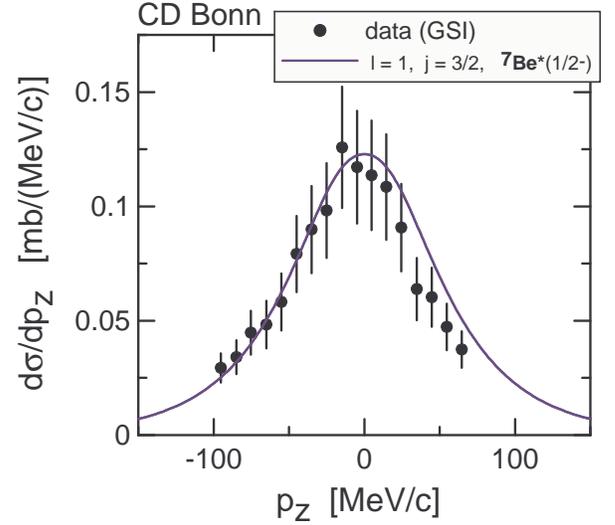}
  \caption{Results for exclusive $p_{z}$ distribution, considering only
the core excited contribution. The solid curve is the calculation
using NCSM wave function with CD-Bonn 2000 interaction.
  \label{GSI_cdb2k_exc}}
\end{figure}

Figure \ref{GSI_cdb2k_exc} shows our results for exclusive $p_{z}$
distribution, considering only the core excited contribution. The
solid curve is the calculation using NCSM wave function with the
CD-Bonn 2000 interaction. The data, from Ref.~\cite{Gil02}, has a
width of $109\pm 7$ MeV/c and the theoretical one is 112 MeV/c. The
total experimental cross section for the excitation of this state is
$\sigma_{-p}=12\pm 3$ mb, whereas the theoretical one is 16.36 mb,
again in good agreement with the data.

\begin{figure}[hbtp]
  \includegraphics*[width=0.9\columnwidth]
   {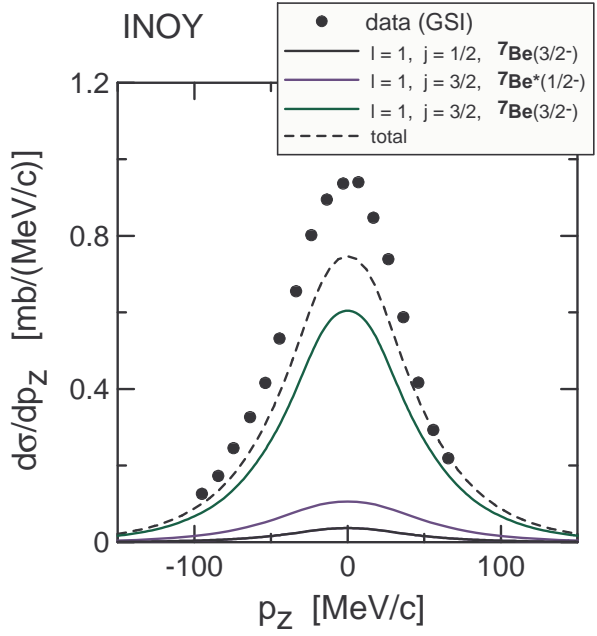}
  \caption{Same as in Fig. \ref{GSI_cdb2k}, but using NCSM wave functions calculated
  with the  INOY interaction.
  \label{GSI_INOY}}
\end{figure}
\begin{figure}[hbtp]
  \includegraphics*[width=0.9\columnwidth]
   {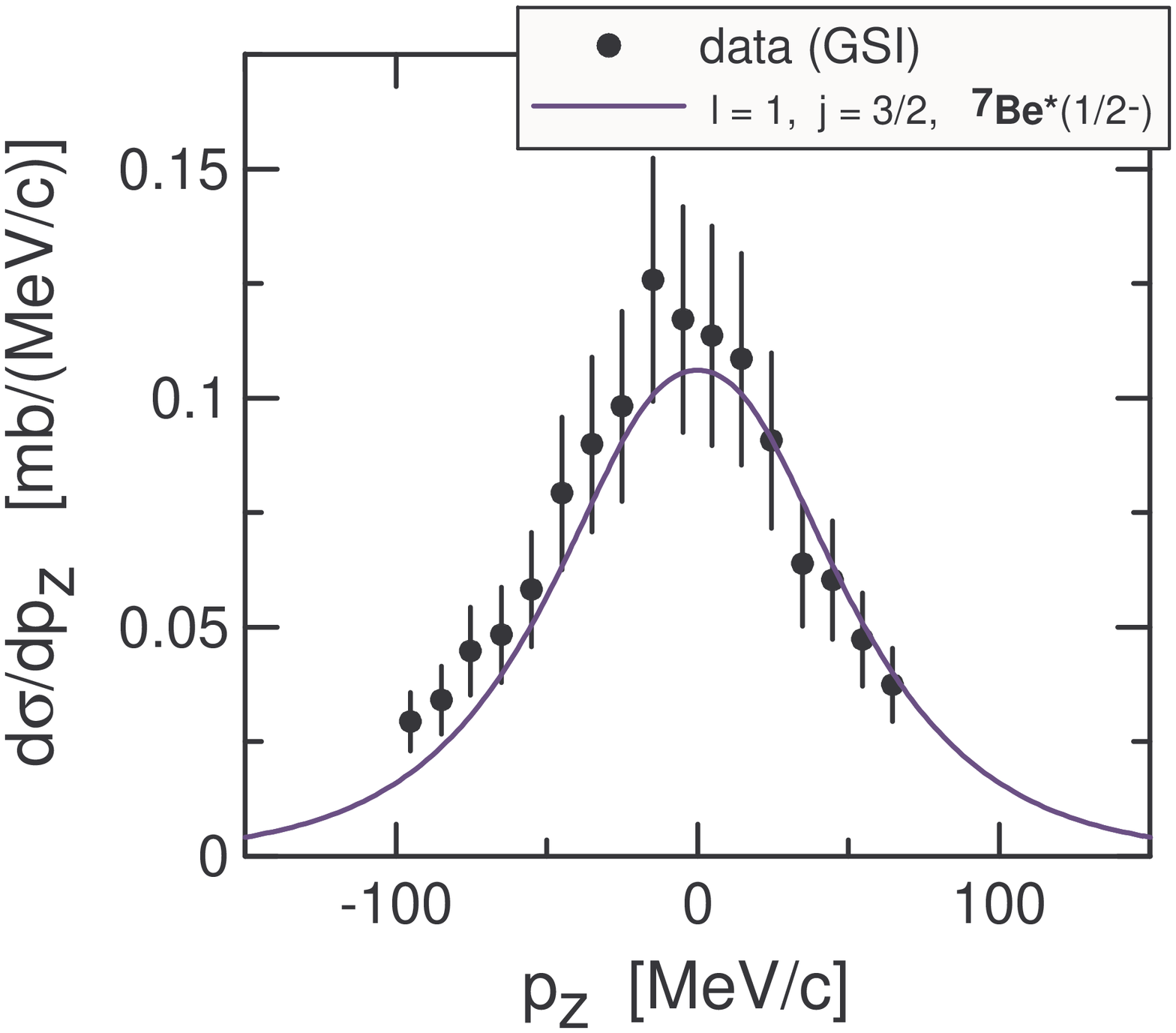}
  \caption{Same as in Fig. \ref{GSI_cdb2k_exc}, but using NCSM excited state wave function calculated
  with the  INOY interaction.
  \label{GSI_INOY_exc}}
\end{figure}

Figures \ref{GSI_INOY} and \ref{GSI_INOY_exc} are the same as the
respective Figs.~\ref{GSI_cdb2k} and \ref{GSI_cdb2k_exc}, but for
NCSM wave functions calculated with the INOY interaction. The
theoretical cross section for the inclusive reaction is 82.93 mb,
and 13.17 mb for the inclusive cross section for the excited state.
These cross sections are smaller than the ones obtained with the
CD-Bonn 2000 interaction. The probable reason is that the INOY
interaction yields a smaller $\langle r^2\rangle_{\rm rms}$-value than the
CD-Bonn 2000 interaction. The calculations yield 92 MeV/c and 108
MeV/c for the for the inclusive and the excited state momentum
distribution widths, respectively. Table \ref{tab:B8cross2} summarizes
the calculations for the total cross sections.

\begin{table}[hbtp]
  \caption{Cross sections for the proton-removal reactions
$^{8}\mathrm{B}+\ ^{9}\mathrm{Be}$ at 41 MeV/nucleon (MSU) and
$^{8}\mathrm{B}+\ ^{12}\mathrm{C}$ at 936 MeV/nucleon (GSI). The
calculated total inclusive cross sections are given by
$\sigma_{\rm inc}^{\left(\mathrm{th-B}\right)}$ for the CD-Bonn 2000
interaction
and $\sigma_{\rm inc}^{\left(\mathrm{th-I}\right)}$ for the INOY
interaction. The $10\hbar\Omega$ model space and HO frequencies of 
12(16) MeV were employed for the CD-Bonn 2000 (INOY).
The calculated cross  sections for the excited state are given by
$\sigma_{\rm exc}^{\left(\mathrm{th-B}\right)}$ and
$\sigma_{\rm exc}^{\left(\mathrm{th-I}\right)}$,
respectively.
  \label{tab:B8cross2}}
  \begin{ruledtabular}
\begin{tabular}
[c]{lllllll}
&  $\sigma_{\rm inc}^{\left(  \mathrm{exp}\right)  }$ & $\sigma_{\rm inc}^{\left(
\mathrm{th-B}\right)  }$ & $\sigma_{\rm inc}^{\left(
\mathrm{th-I}\right)  }$ &   $\sigma_{\rm exc}^{\left(  \mathrm{exp}\right)  }$ &
$\sigma_{\rm exc}^{\left(
\mathrm{th-B}\right)  }$ & $\sigma_{\rm exc}^{\left(
\mathrm{th-I}\right)  }$\\
& [mb] & [mb] & [mb] & [mb] & [mb] & [mb]\\
\hline
MSU  & -- & 82.96 & 71.85 & -- &
15.31 & 13.26\\
GSI & $94\pm 9$ & 99.66 & 82.93 & $12\pm 3$ &
16.36 & 13.17
    \end{tabular}
  \end{ruledtabular}
\end{table}

\section{\label{sec:sfactor}S-factor for $^7{\rm Be(p},\gamma)^8{\rm B}$}

We now turn our focus to the S-factor for the reaction  $^7{\rm
Be(p},\gamma)^8{\rm B}$. This reaction is very important to
understand the structure of our sun. The high energy neutrinos from
the beta-decay of $^8$B come from the sun's center and, therefore, are
a direct measure of the conditions in its interior (temperature,
pressure, chemical composition, etc.). The wave functions calculated
with NCSM method, described in the previous sections will be used
for the purpose.

The main contribution to the S-factor for the radiative capture
reaction  $^7{\rm Be(p},\gamma)^8{\rm B}$ is due to the electric
dipole multipolarity \cite{Radcap}. To calculate the S-factor, we
have to evaluate a matrix element of the electric dipole operator
$\vec{E1}=\sum_{i=1}^A e_i (\vec{r}_i-\vec{R})$ with $e_i=e$ for a proton
and $e_i=0$ for a neutron and the center of mass coordinate 
$\vec{R}=\frac{1}{A}\sum_i \vec{r}_i$. Using the notation introduced in
Eqs. (\ref{proj-targ_state_delta}) and (\ref{cluster_form_factor}),
we calculate a transition from a continuum state 
$| {\cal A}
\Phi_{(l_c\frac{1}{2})j_c;\alpha I_1}^{(A-1,1)J_c};\delta_{r}\rangle
\varphi^E_{l_c j_c \alpha I_1 J_c}(r)$
to the bound state $|A \lambda J_b\rangle$. 
Here, $\varphi^E_{l_c j_c \alpha I_1 J_c}(r)$ is the scattering
wave function of the $^7$Be+p relative motion.
As the internal excitation
of $^7$Be by the electric dipole operator will have a negligible overlap with the
$^8$B ground state, we arrive at the following result
\begin{eqnarray}\label{E1_mat_el}
&& \int_0^\infty dr r^2 \langle A \lambda J_b || E1^{(1)} || {\cal A}
\Phi_{(l_c\frac{1}{2})j_c;\alpha I_1}^{(A-1,1)J_c};\delta_{r}\rangle
\varphi^E_{l_c j_c \alpha I_1 J_c}(r)
\nonumber \\
&& \approx \frac{Z_{A-1}-(A-1)}{A} e \sqrt{\frac{4\pi}{3}} \sqrt{A} 
\nonumber \\
&&\times \int_0^\infty dr r^2 \langle A \lambda J_b || r Y_1 ||
\Phi_{(l_c\frac{1}{2})j_c;\alpha I_1}^{(A-1,1)J_c};\delta_{r}\rangle
\varphi^E_{l_c j_c \alpha I_1 J_c}(r)
\nonumber \\
&& = \frac{Z_{A-1}-(A-1)}{A} e \sum_{l_bj_b} 
\left\{ \begin{array}{ccc} 1 & J_b & J_c \\
  I_1 & j_c & j_b
\end{array}\right\}
\hat{J}_b\hat{J}_c 
\nonumber \\
&&\times\hat{j}_b\hat{j}_c (-1)^{j_b+j_c+I_1+J_c-\textstyle{\frac{1}{2}}}
\left(\begin{array}
[c]{ccc}%
j_c & 1 & j_b\\
-\frac{1}{2} & 0 & \frac{1}{2}%
\end{array}
\right)
\nonumber \\
&& \times \int_0^\infty dr 
g^{A\lambda J_b}_{(l_b\frac{1}{2})j_b;A-1 \alpha I_1}(r)
r^3 \varphi^E_{l_c j_c \alpha I_1 J_c}(r) \; ,
\end{eqnarray}
with $|A \lambda J_b\rangle$ the ground state of $^8$B with $J_b^\pi=2^+$ and
$|A-1 \alpha I_1\rangle$ the ground state of $^7$Be with $I_1^\pi=\frac{3}{2}^-$.
The $^7$Be proton number $Z_{A-1}$ is equal to 4. The overlap integral
$g^{A\lambda J_b}_{(l_b\frac{1}{2})j_b;A-1 \alpha I_1}(r)$ is defined
as in Eq. (\ref{cluster_form_factor}) with its asymptotic tail corrected
as described in Sect.~\ref{subsec:corr_overlap}. 
Assuming that the relative-motion scattering wave function is independent 
on the continuous state total angular momentum $J_c$, 
$\varphi^E_{l_c j_c \alpha I_1 J_c}(r)=\varphi^E_{l_c j_c \alpha I_1}(r)$,
we arrive at the S-factor expression
\cite{Robertson,Radcap}
\begin{eqnarray}
S(E) &=&{\frac{4\pi ^{2}e_{\rm eff}^2\hbar^2}{3\mu_{{\rm p}^7{\rm Be}}}}
\left( {\frac{{E+E_b}}{{%
\hbar c}}}\right) ^{3}\exp {\left[ 2\pi \eta
(E)\right]}
\nonumber \\
&\times&  {\frac{(2J_b+1)}{2(2I_1+1)}} \sum_{l_b j_b l_c j_c}
\left(
\begin{array}
[c]{ccc}%
j_c & 1 & j_b\\
-\frac{1}{2} & 0 & \frac{1}{2}%
\end{array}
\right)^2\nonumber \\
&\times & |\int_0^\infty dr 
g^{A\lambda J_b}_{(l_b\frac{1}{2})j_b;A-1 \alpha I_1}(r)
r^3 \varphi^E_{l_c j_c \alpha I_1}(r)|^{2}
\ , \label{Sfact}
\end{eqnarray}%
where $E_b$ is the binding energy of the $^8$B bound state with respect to
$^7$Be+p.
Further, $\eta(E)=Z_aZ_Ae^2/\hbar v_{{\rm p}^7{\rm Be}}$ is
the Sommerfeld parameter, $k$ is the relative momentum, 
$E=\hbar^2 k/2\mu_{{\rm p}^7{\rm Be}}$ is the center-of-mass energy of the proton+$^7$Be,
and $e_{\rm eff}=\frac{(A-1)-Z_{A-1}}{A}e=3e/8$ is the effective charge for 
the E1 capture for this specific reaction. The continuum wave functions 
are normalized so that
\begin{eqnarray}
&&r\varphi^E_{l_c j_c \alpha I_1}(r\rightarrow \infty) \longrightarrow
\nonumber \\
&& i\sqrt{\frac{\mu_{{\rm p}^7{\rm Be}}}{2\pi k \hbar^2}}\left[ H_{l_c}^{(-)}(r) -{\cal
S}_{l_cj_c}H_{l_c}^{(+)}(r)\right]e^{i\sigma_{l_c}(E)},
\label{contnorm}
\end{eqnarray}
where ${\cal S}_{l_cj_c}=\exp{\left[2i\delta_{l_cj_c}\right]}$, with $\delta_{l_cj_c}$
($\sigma_{l_c}$) being the nuclear (Coulomb)
phase shift, and $H^{(\pm)}_{l_c}(r) = G_{l_c}(r)\pm iF_{l_c}(r)$, with $F_{l_c}$
and $G_{l_c}$ being the regular and irregular Coulomb wave functions, respectively.
The normalization (\ref{contnorm}) implies that the continuum wavefunctions
satisfy the relation
$$
\left\langle \varphi_{\eta}^{E}|\varphi_{\eta^\prime}^{E^\prime}
\right\rangle =\delta\left(  E-E^{\prime}\right)  \delta_{\eta \eta^{\prime}}
.\label{cont_norm}%
$$

Eq. (\ref{Sfact}) shows that the S-factor is a more stringent
test of the nuclear model. Whereas the momentum distributions
discussed in the previous section depends only on the bound state
overlap function, $g^{A\lambda J_b}_{(l_b\frac{1}{2})j_b;A-1 \alpha I_1}(r)$, 
obtained from the corrected NCSM overlap integrals
discussed in subsection \ref{subsec:corr_overlap},
the S-factor also depends on the continuum wave function,
$\varphi^E_{l_c j_c \alpha I_1}(r)$. Presently, we have not yet developed 
a theory to extend the NCSM in order to describe continuum wave functions.
Therefore, to obtain  $\varphi^E_{l_c j_c \alpha I_1}(r)$ for $s$ and $d$ waves,
we use a potential model with a Woods-Saxon+Coulomb+spin-orbit interaction. 
In this respect, our approach is similar to that by Nollett {\it et al.} 
\cite{Nollett_ad,Nollett_ag} where {\it ab initio} variational Monte Carlo
wave functions were used in combination with cluster-cluster potential
model scattering wave functions to describe d($\alpha,\gamma$)$^6$Li,
$^3$H($\alpha,\gamma$)$^7$Li and $^3$He($\alpha,\gamma$)$^7$Be capture
reactions. Since the largest part
of the integrand of the last term in Eq. \ref{Sfact} stays outside the
nuclear interior, one expects that the continuum wave functions are
well described by this model. It is possible to use for this purpose
the same WS potential that we obtained from correcting the bound-state
overlap integral. Then we would have a different scattering state for each
bound-state partial wave and each NCSM model space and HO frequency.
In order to have the same scattering wave function in all the calculations,
we chose a WS potential from Ref.~\cite{Esbensen} that was fitted to
reproduce the $p$-wave $1^+$ resonance in $^8$B. It was argued \cite{Robertson}
that such a potential is also suitable for the description of $s$- and $d$-waves.
The parameters of this potential are summarized in Table~\ref{tab:WSparam}.

We note that the S-factor results are very weakly dependent on the choice
of the potential that describes the scattering state. This is in particular
true at low energies. Using our fitted potentials for the scattering state 
instead of the scattering potential from Table~\ref{tab:WSparam}
changes (typically increases) the S-factor 
by less than 1.5 eV b at 1.6 MeV with still a smaller change at lower energies
and no change at 0 MeV.

At the same time, we note that the $^7$Be+p scattering length has been measured
\cite{p7Be_scl}. A potential model with no spin-orbit term was developed in 
Ref.~\cite{Davids03} that fits the experimental scattering length. It is not
consistent to use that potential in the present work, as we employ an alternative
angular momentum coupling scheme. Still, to get a further insight on the
S-factor energy dependence sensitivity to the scattering wave function, we 
performed test calculations with the $s=2$ potential of Ref.~\cite{Davids03}.
Up to 100 keV, we observed very little change of the S-factor compared
to the situation when the scattering potential from Table~\ref{tab:WSparam}
was used. At higher energies, the S-factor increased by up to 5 eV b with a difference
of 4 eV b at 1.6 MeV. To resolve the issue of the S-factor energy dependence at
higher energies, we need to develop a theory that extends the NCSM to describe 
scattering states. This is a subject of current and future investigations. 

\begin{figure}[hbtp]
  \includegraphics*[width=0.9\columnwidth]
   {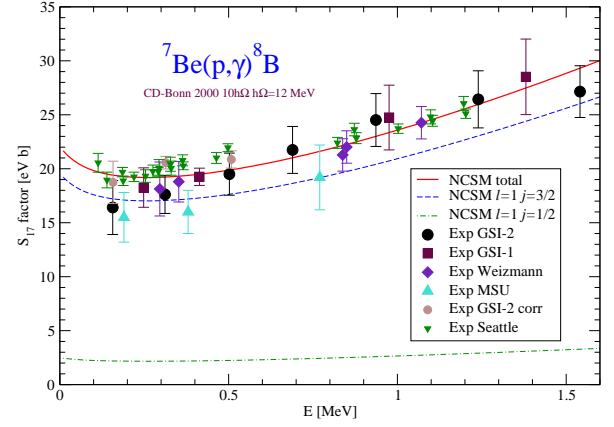}
  \caption{The $^7$Be(p,$\gamma$)$^8$B S-factor obtained
using the NCSM cluster form factors with corrected asymptotics
by the WS solution fit.
The dashed and dashed-dotted lines show the contribution due to the $l=1$,
$j=3/2$ and $j=1/2$ partial waves, respectively.
The CD-Bonn 2000 NN potential, the $10\hbar\Omega$ model space
and the HO frequency of $\hbar\Omega=12$ MeV were used.
Experimental values are from Refs. \protect\cite{Baby,Seattle,Be7pgamm_exp}.
  \label{S-factor_12_partial}}
\end{figure}

In Fig.~\ref{S-factor_12_partial}, we present
our first application for the astrophysical S-factor of 
the $^7{\rm Be(p},\gamma)^8{\rm B}$ reaction. 
We use bound-state wave functions
calculated with the CD-Bonn 2000 interaction in the $10\hbar\Omega$ model
space and the optimal HO frequency $\hbar\Omega=12$ MeV.
The WS solution fit procedure was employed
to correct the asymptotics of the NCSM overlap functions.
In the figure, we show the S-factor contributions from the dominant
$l=1$, $j=3/2$ and $j=1/2$ partial waves by the dashed lines. Clearly, the
$j=3/2$ partial wave is the most important one.
The upper curve, the full line, is the sum of the two contributions.
The experimental data is a compilation of the latest experiments for the
S-factor. They include direct, as well as some indirect measurements
(Coulomb dissociation).

The slope of the curve corresponding to the total S-factor follows
the trend of the data.  
From Fig. \ref{S-factor_12_partial} it is also clear
that the spectroscopic factors for the $l=1$, $j=3/2$ and $j=1/2$
partial waves are well  described within the NCSM and the CD-Bonn
2000 interaction. Our calculation presented in Fig.~\ref{S-factor_12_partial}
is in a very good agreement with the recent direct measurement data
of Ref.~\cite{Seattle}. We note that our S-factor energy dependence
more resembles that obtained within the three-cluster model of Ref.~\cite{D04}
rather than that obtained within the potential model of Ref.~\cite{Davids03}.  

\begin{figure}[hbtp]
  \includegraphics*[width=0.9\columnwidth]
   {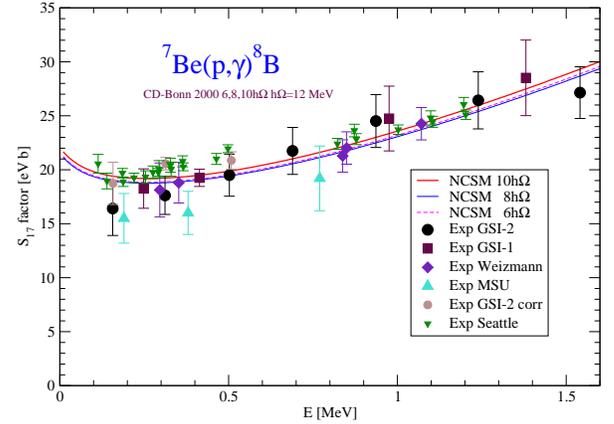}
  \caption{The $^7$Be(p,$\gamma$)$^8$B S-factor obtained
using the NCSM cluster form factors with corrected asymptotics
by the WS solution fit.
The dependence on the size of the basis from $6\hbar\Omega$ to $10\hbar\Omega$
is shown. The CD-Bonn 2000 NN potential and the HO frequency of $\hbar\Omega=12$
MeV were used.
Experimental values are from Refs. \protect\cite{Baby,Seattle,Be7pgamm_exp}.
  \label{S-factor_12_Nmax}}
\end{figure}

In order to judge the convergence of our S-factor calculation, we performed
a detailed investigation of the model-space-size and the HO frequency dependencies.
The dependence of the  $^7$Be(p,$\gamma$)$^8$B S-factor on the size
of the basis from $6\hbar\Omega$ to $10\hbar\Omega$ corresponding to the above
discussed calculation using the CD-Bonn 2000 and the HO frequency
of $\hbar\Omega=12$  MeV is shown in Fig. \ref{S-factor_12_Nmax}.
One observes a very little model-space dependence with some oscillatory behavior.
The fact that the calculation converges for a basis size within this
energy interval demonstrates that the optimal HO frequency determined from the
ground-state energy calculation is also close to optimal for the S-factor calculation.

\begin{figure}[hbtp]
  \includegraphics*[width=0.9\columnwidth]
   {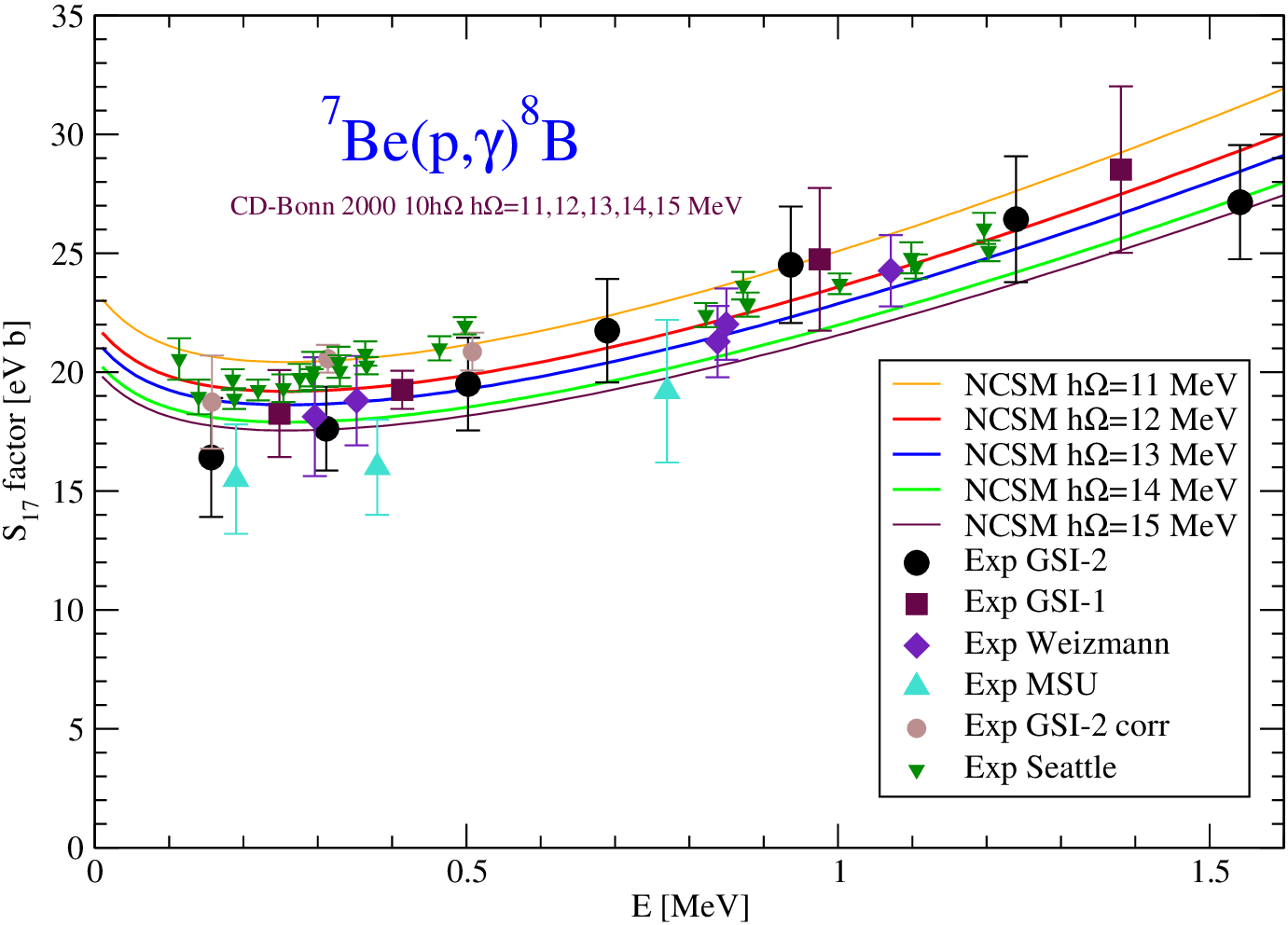}
  \caption{The $^7$Be(p,$\gamma$)$^8$B S-factor obtained
using the NCSM cluster form factors with corrected asymptotics
by the WS solution fit.
The dependence on the size of the HO frequency from $\hbar\Omega=11$ MeV
to $\hbar\Omega=15$ MeV is shown.
The CD-Bonn 2000 NN potential and the $10\hbar\Omega$ model space were used.
Experimental values are from Refs. \protect\cite{Baby,Seattle,Be7pgamm_exp}.
  \label{S-factor_11_15_freq}}
\end{figure}

\begin{figure}[hbtp]
  \includegraphics*[width=0.9\columnwidth]
   {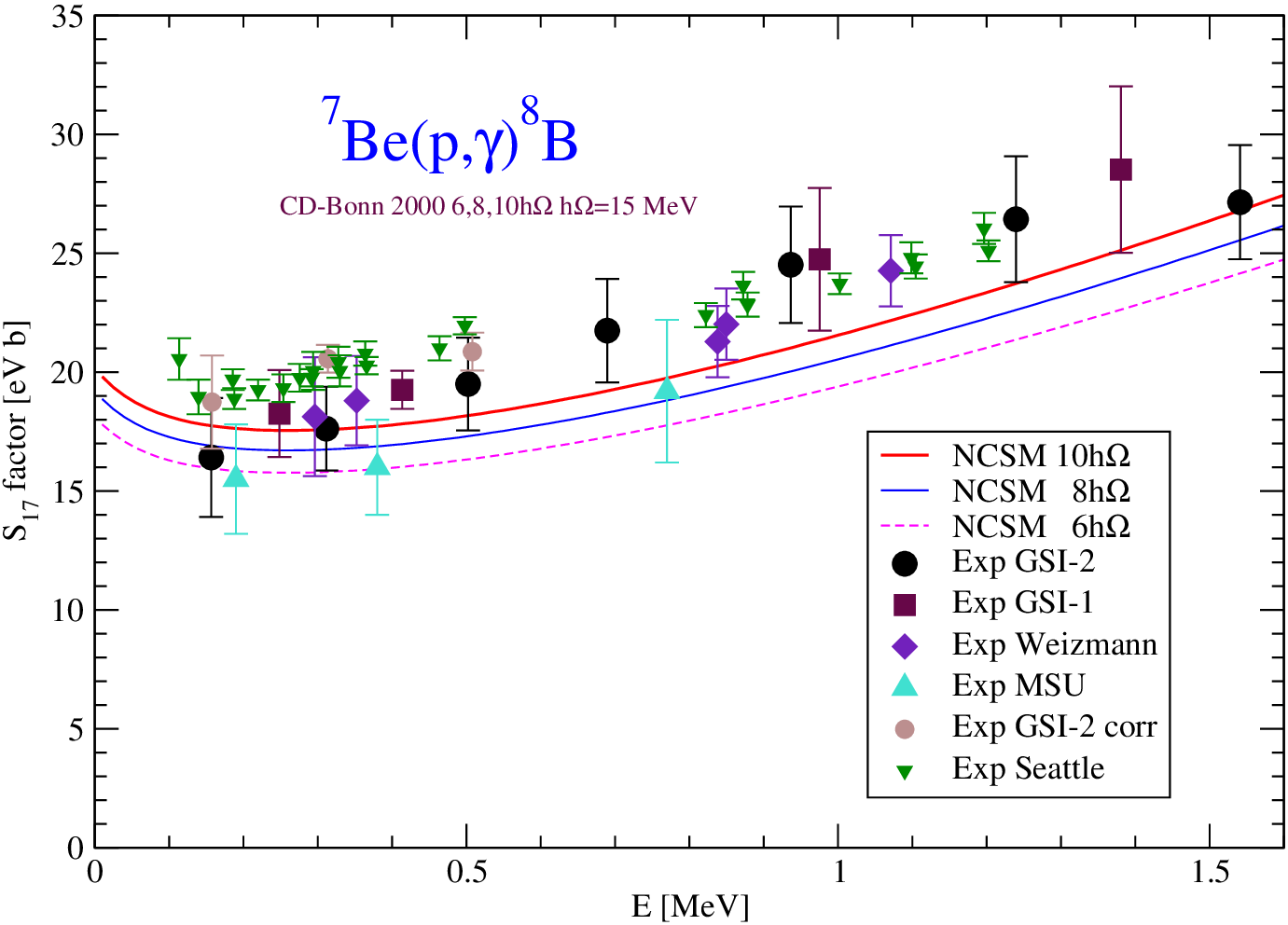}
  \caption{Same as in Fig.~\protect\ref{S-factor_12_Nmax} but for
the HO frequency of $\hbar\Omega=15$ MeV.
  \label{S-factor_15_Nmax}}
\end{figure}

\begin{figure}[hbtp]
  \includegraphics*[width=0.9\columnwidth]
   {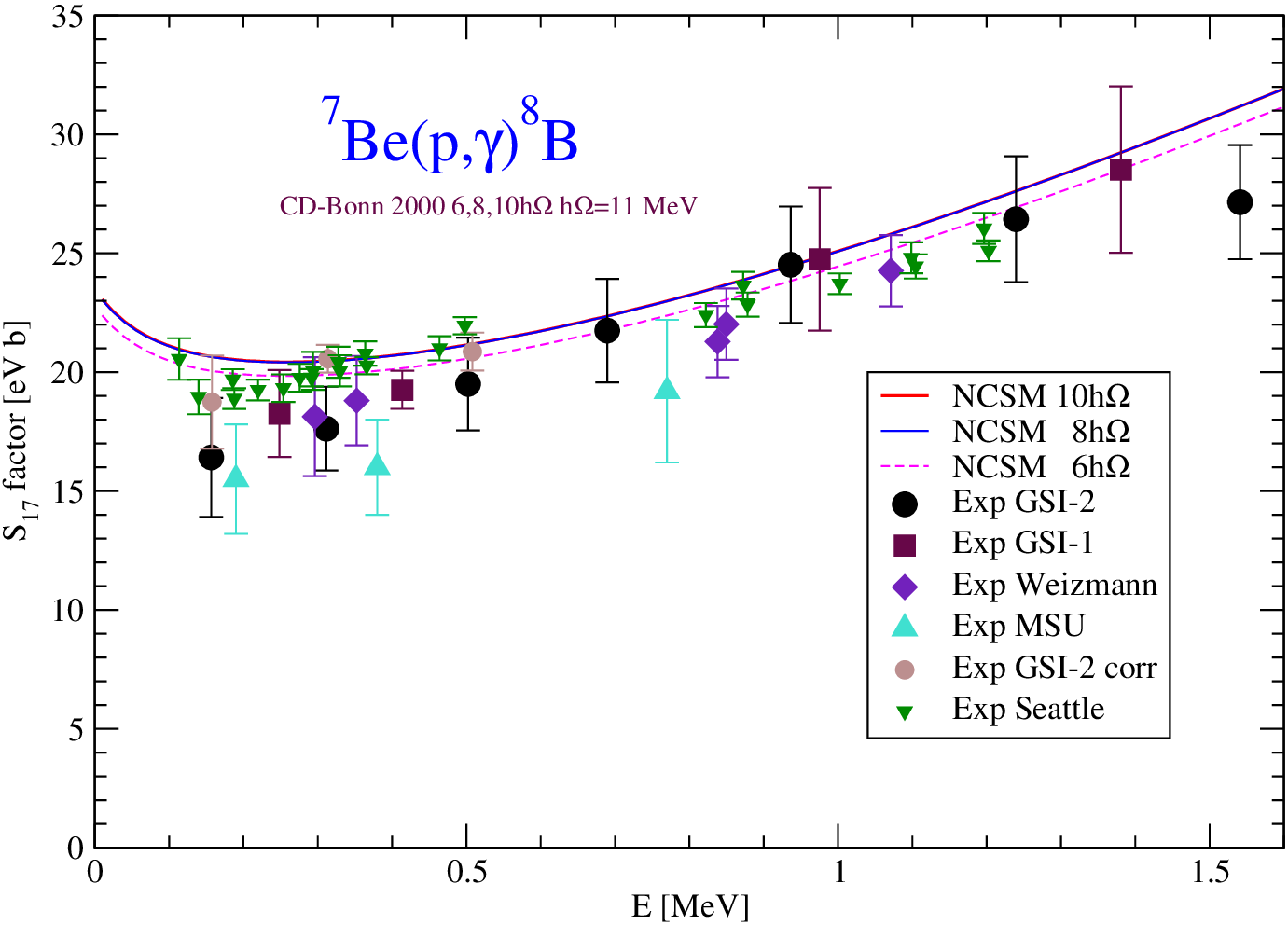}
  \caption{Same as in Fig.~\protect\ref{S-factor_12_Nmax} but for
the HO frequency of $\hbar\Omega=11$ MeV.
  \label{S-factor_11_Nmax}}
\end{figure}

To further investigate the HO frequency sensitivity, we show in
Fig. \ref{S-factor_11_15_freq} the dependence of the
$^7$Be(p,$\gamma$)$^8$B S-factor on the HO frequency from
$\hbar\Omega=11$ MeV  to $\hbar\Omega=15$ MeV. The CD-Bonn 2000 NN
potential and the $10\hbar\Omega$ model space were used. 
Again, the WS solution fit procedure was employed
to correct the asymptotics of the NCSM overlap functions.
In all
cases, we obtain basically identical energy dependence. The absolute
values of the S-factor increase with decreasing frequency. To
determine the optimal frequency and interpolate the converged
S-factor result, we examine the basis size dependence for different
HO frequencies. In Fig.~\ref{S-factor_15_Nmax}, we show the
$\hbar\Omega=15$ MeV results for model spaces from $6\hbar\Omega$ to
$10\hbar\Omega$. We observe a steady increase of the S-factor with
the basis size enlargement. Contrary to this situation, the
calculation using the HO frequency of $\hbar\Omega=11$ MeV presented
in Fig.~\ref{S-factor_11_Nmax} shows that the S-factor does not
increase any more with increasing $N_{\rm max}$. Actually, there is
a small decrease when going from the $8\hbar\Omega$ to the
$10\hbar\Omega$. In Table~\ref{tab:S-factor}, we summarize our
S-factors at 10 keV obtained for different frequencies and model
spaces. 
In addition to the results obtained using the WS solution fit procedure, 
we also present the $S_{17}$ and the ANC obtained using the alternative
direct Whittaker matching procedure. 
We note that the ANC from our {\it ab initio} approach 
are smaller than those obtained
within the microscopic three-cluster model \cite{D04} but still larger
than the experimental ones from the DWBA analysis of Ref.~\cite{Azhari}.
In general, both procedures lead to basically
identical energy dependence with a difference of about 1 to 2 eV b in the S-factor
with the smaller values from the direct Whittaker function matching procedure.
Taking into account that in the case of the direct Whittaker function matching
the $S_{17}$ increases with $N_{\rm max}$ even at
the HO frequency of $\hbar\Omega=11$ MeV, unlike in the case of the WS solution 
fit procedure, results of the two approaches do not contradict each other.
Combining all these results, we determine that the optimal
frequency is between $\hbar\Omega=11$ and 12 MeV.
Results in this frequency region show very weak dependence on $N_{\rm max}$, 
with relative difference between the two methods always in the range of 5 to 8\%.
The full range of results is covered by  $S_{17}(10\;{\rm keV})=22.1\pm 1.0$ eV b.


\begin{figure}[hbtp]
  \includegraphics*[width=0.9\columnwidth]
   {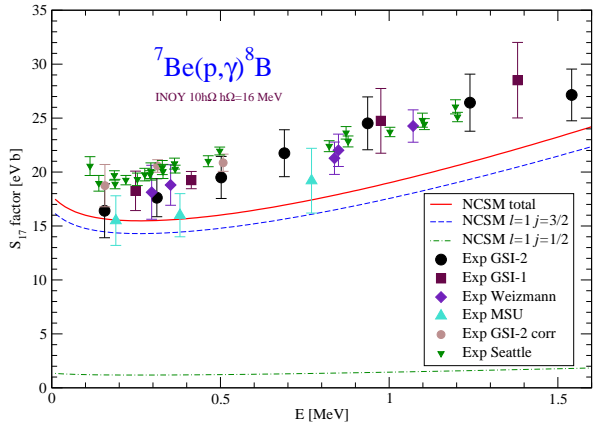}
  \caption{Same as in Fig.~\protect\ref{S-factor_12_partial} but for
the INOY NN potential and the HO frequency of $\hbar\Omega=16$ MeV.
  \label{S-factor_INOY_16_partial}}
\end{figure}

\begin{figure}[hbtp]
  \includegraphics*[width=0.9\columnwidth]
   {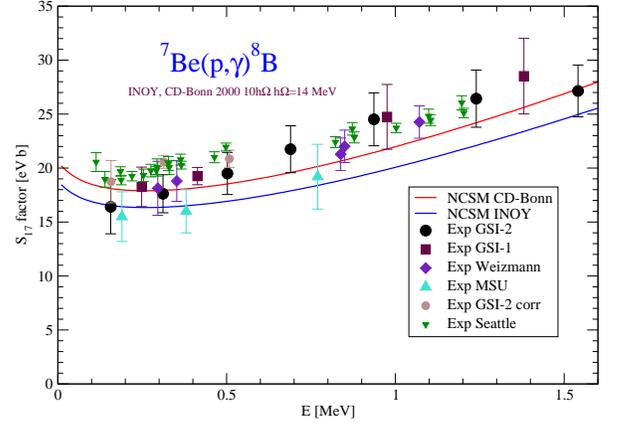}
  \caption{The $^7$Be(p,$\gamma$)$^8$B S-factor obtained
using the NCSM cluster form factors with corrected asymptotics 
by the WS solution fit. The result obtained using the CD-Bonn and
INOY NN potentials are compared. The $10\hbar\Omega$ model space and
HO frequency of $\hbar\Omega=14$ MeV were used. Experimental values
are from Refs. \protect\cite{Baby,Seattle,Be7pgamm_exp}.
  \label{S-factor_CDB_INOY_14}}
\end{figure}

We have performed a similar, although a less extensive study for
the INOY NN interaction. In Fig.~\ref{S-factor_INOY_16_partial}, we present
the S-factor and its partial wave contributions as obtained using the INOY NN
interaction and the ground-state-energy-determined optimal frequency
of $\hbar\Omega=16$ MeV. The $10\hbar\Omega$ model space 
and the WS solution fit procedure was utilized. The energy
dependence is slightly weaker compared to the CD-Bonn case and the S-factor underestimates
most of the experimental data. To make a direct comparison with the CD-Bonn calculation,
in Fig.~\ref{S-factor_CDB_INOY_14} we show S-factors obtained from the CD-Bonn 2000 and the INOY NN
potentials using identical HO frequency of $\hbar\Omega=14$ MeV and the $10\hbar\Omega$
model space. Although the spectroscopic factors are almost the same, see Table~\ref{tab:B8specfac},
the INOY S-factor is significantly smaller than that of the CD-Bonn 2000. This is a consequence
of different shapes of the overlap functions displayed in Fig.~\ref{B8_Be7+p_overlap_cdb2k_INOY_14}.
This result is correlated with our radius and quadrupole moment results as well as results
of momentum distributions in knockout reactions.
The fact that the INOY NN interaction yields a smaller matter radii than the
CD-Bonn 2000 leads to a smaller S-factor.
The S-factor convergence for the INOY NN potential is similar or better than for the CD-Bonn 2000.
Based on the results presented in the figures and the Table~\ref{tab:S-factor}, we estimate
the INOY S-factor result at 10 keV to be $19.0\pm 1$ eV b.

\begin{table}[hbtp]
  \caption{The calculated $^7$Be(p,$\gamma$)$^8$B S-factor, in eV b, at the energy of 10 keV.
Two ways of correcting the NCSM overlap asymptotics, by the Woods-Saxon potential solution fit (WS)
and by a direct Whittaker function matching (Whitt), are compared. The asymptotic normalization 
constants, in fm$^{-1/2}$, correspond to the Whittaker function matching case.    
Results obtained using the CD-Bonn 2000 and INOY NN potentials at different
HO frequencies and model spaces as well as the NCSM extrapolated $S_{17}$ values with their estimated
errors are presented.
  \label{tab:S-factor}}
  \begin{ruledtabular}
    \begin{tabular}{cccccc}
\multicolumn{6}{c}{CD-Bonn 2000} \\
$\hbar\Omega$ [MeV] & $N_{\rm max}$ & $C_{1,3/2}$ & $C_{1,1/2}$ & $S_{17}^{\rm Whitt}$ & $S_{17}^{\rm WS}$ \\
\hline
15 & 6  & 0.647 & 0.195 & 16.81 & 17.80 \\
15 & 8  & 0.660 & 0.206 & 17.58 & 18.87 \\
15 & 10 & 0.672 & 0.216 & 18.33 & 19.81 \\
14 & 10 & 0.680 & 0.220 & 18.78 & 20.21 \\
13 & 10 & 0.692 & 0.234 & 19.64 & 21.02 \\
12 & 6  & 0.693 & 0.240 & 19.75 & 21.24 \\
12 & 8  & 0.696 & 0.242 & 19.96 & 21.14 \\
12 & 10 & 0.704 & 0.247 & 20.45 & 21.66 \\
11 & 6  & 0.715 & 0.261 & 21.30 & 22.38 \\
11 & 8  & 0.715 & 0.263 & 21.33 & 23.04 \\
11 & 10 & 0.720 & 0.262 & 21.60 & 23.06 \\
\multicolumn{4}{c}{NCSM $S_{17}(10\;{\rm keV})$} & \multicolumn{2}{c}{$22.1\pm 1.0$} \\
\multicolumn{6}{c}{INOY} \\
$\hbar\Omega$ [MeV] & $N_{\rm max}$ & $C_{1,3/2}$ & $C_{1,1/2}$ & $S_{17}^{\rm Whitt}$ & $S_{17}^{\rm WS}$ \\
\hline
16 & 10 & 0.641 & 0.182 & 16.34 & 17.49 \\
15 & 10 & 0.649 & 0.189 & 16.83 & 17.95 \\
14 & 6  & 0.652 & 0.190 & 16.94 & 17.78 \\
14 & 8  & 0.654 & 0.194 & 17.12 & 18.21 \\
14 & 10 & 0.660 & 0.198 & 17.44 & 18.46 \\
\multicolumn{4}{c}{NCSM $S_{17}(10\;{\rm keV})$} & \multicolumn{2}{c}{$19.0\pm 1.0$} \\
    \end{tabular}
  \end{ruledtabular}
\end{table}
%

%
\section{\label{sec:conc}Conclusions}
We studied nuclear structure of $^7$Be, $^8$B and $^{7,8}$Li within
the {\it ab initio} NCSM. We calculated overlap integrals of the $^8$B ground
state with $^7$Be+p as a function of the separation between the proton and $^7$Be
using NCSM wave functions obtained in model spaces up to $10\hbar\Omega$. Assuming
that the NCSM overlap integrals are realistic in the interior part, we utilized
Woods-Saxon potential as a tool to correct their asymptotics.
We performed a least-square fit of WS potential solutions
to the NCSM overlap integrals in the range from 0 fm up to about 4 fm under the
constraint that the experimental separation energy is reproduced.
In addition, we employed an alternative procedure of a direct Whittaker function matching
to correct the asymptotic tail of the NCSM overlap integrals.
The corrected overlap integrals were then used for the
$^7$Be(p,$\gamma$)$^8$B S-factor calculation. We investigated the dependence of the
S-factor on the size of the model space and on the HO frequency as well as on the procedure
of the asymptotic tail correction. Based on this investigation,
we arrived at the S-factor result at 10 keV to be $22.1\pm 1.0$ eV b
and $19.0\pm 1.0$ eV b for the CD-Bonn 2000 and INOY NN potentials, respectively.
We note that the CD-Bonn 2000 results for the point-proton rms radii, quadrupole moments 
and momentum distributions in knockout reactions with $^8$B projectiles
are in a better agreement with experiment than those obtained using the INOY NN potential.
At the same time, the CD-Bonn 2000 predicts the point-proton rms radii of $^4$He and $^6$He
in agreement with experiment, while the INOY NN potential underpredicts both \cite{He6_rad}.
Therefore, we consider the CD-Bonn 2000 S-factor result more realistic. The energy dependence
as well as the zero energy value of the S-factor calculated using the CD-Bonn 2000
NN potential are in a very good agreement with recent direct measurement data
of Ref.~\cite{Seattle}. We stress that no adjustable parameters were used in our
{\it ab initio} calculations of the $^8$B and $^7$Be bound states. Taking into account
that the S-factor is only weakly dependent on the potential model used to obtain
the scattering state, we consider our results as the first {\it ab initio} prediction
of the $^7$Be(p,$\gamma$)$^8$B S-factor, in particular of its normalization.

%
\begin{acknowledgments}
We thank W. E. Ormand, C. Forss\'en and J. Dobe\v{s} for useful discussions.
This work was partly performed under the auspices of the
U. S. Department of Energy by the University of California, Lawrence
Livermore National Laboratory under contract No. W-7405-Eng-48. Support
from the LDRD contract No.~04--ERD--058, and from U.S.~Department of
Energy, Office of Science, (Work Proposal Number SCW0498),
and  grant No. DE-FG02-04ER41338, is acknowledged.
\end{acknowledgments}

\end{document}